\documentclass[a4paper,11pt]{article}

\usepackage{jheppub} 
\usepackage{hyperref}
\usepackage[utf8]{inputenc}
\usepackage{physics}
\usepackage{amsmath, amsfonts, amsthm, amssymb}
\usepackage{bm} 
\usepackage{mathtools}
\usepackage{lipsum}
\usepackage{ragged2e}
\usepackage{mathrsfs}
\usepackage{xcolor}
\usepackage{youngtab}
\usepackage{scalerel}
 
\DeclareMathAlphabet\mathchorus{T1}{qzc} {m} {n}

\def\Re{\mathrm{Re}\hskip1pt}
\def\Im{\mathrm{Im}\hskip1pt}

\def\eg{{\it e.g.}}
\def\ie{{\it i.e.}}

\def\so{{\mathfrak{so}}}

\def\epsilonzero{\overset{\scriptscriptstyle 0}\varepsilon}
\def\omegazero{\overset{\scriptscriptstyle 0}\omega}
\def\Omegazero{\overset{\scriptscriptstyle 0}\Omega}

\def\Rzero{\overset{\scriptscriptstyle 0}R}

\def\nn{\nonumber}

\def\so{\mathfrak{so}}

\def\*{\partial}

\def\transpose{\intercal}

\def\sF{{\mathscr F}}

\def\RR{{\mathbb R}}
\def\HH{{\mathbb H}}
\def\ZZ{{\mathbb Z}}

\def\Re{\mathrm{Re}\hskip1pt}
\def\Im{\mathrm{Im}\hskip1pt}

\def\su{\mathfrak{su}}

\def\II{{\mathscr I}}

\def\Vol{\hbox{Vol}}

\newcommand{\paral}{\mathbin{\!/\mkern-5mu/\!}}

\def\Ppar{P_{\scaleto{\paral}{6pt}}}
\def\Pperp{P_{\scaleto{\perp}{5pt}}}

\title{Curvature of an exotic 7-sphere}

\author[a]{David S. Berman,}
\author[b]{Martin Cederwall}
\author[a]{and Tancredi Schettini Gherardini}

\affiliation[a]{Centre for Theoretical Physics, School of Physical and Chemical Sciences,
Queen Mary University of London, 327 Mile End Road, London E1 4NS, UK}
\affiliation[b]{Department of Physics, Chalmers Univ. of Technology,
SE-412 96 Gothenburg, Sweden}

\emailAdd{d.s.berman@qmul.ac.uk}
\emailAdd{martin.cederwall@chalmers.se}
\emailAdd{t.schettinigherardini@qmul.ac.uk}

\abstract{We study the geometry of the Gromoll--Meyer sphere, one of Milnor's exotic $7$-spheres. We focus on a Kaluza--Klein Ansatz, with a round $S^4$ as base space, unit $S^3$ as fibre, and $k=1,2$ $SU(2)$ instantons as gauge fields, where all quantities admit an elegant description in quaternionic language. The metric's moduli space coincides with the $k=1,2$ instantons' moduli space quotiented by the isometry of the base, plus an additional $\mathbb{R}^+$ factor corresponding to the radius of the base, $r$. 
We identify a ``center'' of the $k=2$ instanton moduli space with enhanced symmetry.
This $k=2$ solution is used together with the maximally symmetric $k=1$ solution to obtain a metric of maximal isometry, $SO(3)\times O(2)$, and to explicitly compute its Ricci tensor. This allows us to put a bound on $r$ to ensure positive Ricci curvature, which implies various energy conditions for an $8$-dimensional static space-time. This construction then enables a concrete examination of the properties of the sectional curvature.}

\makeatletter
\gdef\@fpheader{\begin{flushright} QMUL-PH-24-22
\end{flushright}
\vspace{-3cm}}
\makeatother

\begin{document}

\frenchspacing

\maketitle
\flushbottom

\section{Introduction}
Exotic spheres in seven dimensions (from now on simply referred as \textit{exotic spheres}) are smooth manifolds homeomorphic to the topological 7-sphere, but not diffeomorphic to the smooth 7-sphere with any ``standard" atlas (such as the one obtained by stereographic projection from the two poles, for instance). Their construction by Milnor in \cite{10.2307/1969983} proved for the first time the existence of pairs of manifolds $(S,\Sigma)$ with these properties, \ie, homeomorphic but not diffeomorphic. The ``non-standard'' member $\Sigma$ of such a pair is known as an \textit{exotic manifold}. Exotic spheres are seven-dimensional compact manifolds with a unique spin structure, which makes them suitable candidates for compactification of M-theory to four dimensions. Their appearance in this context has been put forward (\cite{YAMAGISHI198447}, \cite{FREUND1985263}) but never concretely realised; not in eleven-dimensional supergravity, nor in any other higher-dimensional theory.\footnote{Exotic spheres also appeared in other areas of theoretical physics. For instance, they were considered in the context of gravitational instantons in  \cite{Witten:1985xe} and \cite{10.1063/1.529078}. They were studied within general relativity in \cite{Asselmeyer:1996bh}. Finally, recent studies of Riemannian and Lorentzian metrics on exotic spheres can be found in \cite{cavenaghi2024hearingexoticsmoothstructures} and \cite{CAVENAGHI2024102121}.}

The geometry of exotic spheres has been thoroughly studied in the mathematical literature, and a number of theorems about the existence of metrics with specific properties have been proven (see \cite{nuimeprn10073} for a review). Out of the 27 exotic spheres, 15 of them can be obtained as associated $S^3$ bundles over $S^4$, as they appear in Milnor's original paper; we focus on a specific member of this family, known as the Gromoll--Meyer sphere (\cite{10.2307/1971078}). By construction, the metric on it can be obtained by ``twisting" the product metric on $S^4 \times S^3$ by a connection defined on the bundle. The procedure of constructing a higher-dimensional metric from lower-dimensional geometrical constituents of the bundle is sometimes referred as \textit{inverse Kaluza--Klein} approach (since it is the opposite logic to the Kaluza--Klein dimensional reduction for field theories). The resulting metric is known as \textit{Kaluza--Klein metric}. In this paper, we investigate the family of Kaluza--Klein metrics on the Gromoll--Meyer sphere, by performing very explicit calculations to obtain some of its main properties. We compare our findings to the existing results in the mathematical literature and comment on their physical implications in formulating static solutions of general relativity in eight dimensions. Our detailed results could be taken as a starting point for a number of further investigations within the context of supergravity, but also for a careful mathematical investigation of the sectional curvature of the metric on the Milnor bundle.

We aim for the paper to be as self-contained as possible. Therefore, we put some effort into introducing the notation and rederiving well known results before applying it to the actual construction of exotic spheres.
The paper is structured as follows. In Section \ref{sec:Quat_inst_and_spheres}, we introduce the quaternionic notation adopted in this paper, and illustrate how all of the geometrical quantities that appear in the Kaluza--Klein metric admit a natural description in terms of quaternionic-valued objects. In Section \ref{sec:Curvature}, we derive the general expression for the Ricci curvature and Ricci scalar associated to the Kaluza--Klein Ansatz. Section \ref{sec:KKModuli_vs_Instanton_moduli} is devoted to an explicit construction of the $k=1$ and $k=2$ $SU(2)$ instantons, mainly focussing on their moduli space and on how to switch between the singular/regular gauge expressions; we also discuss the relation between the instantons' moduli space and the Kaluza--Klein metric's moduli space. This analysis motivates a special choice of moduli for the $k=1$ and $k=2$ instantons, assumed throughout Section \ref{CalcRicciSec}, where we show that the corresponding Kaluza--Klein metric has maximal isometry, \ie, $SO(3)\times O(2)$ (\cite{10.2307/1971078}). Moreover, the Ricci tensor is explicitly computed and a condition for it to be positive is found. In Section \ref{sec:Energy_conditions}, this result is used to assess the energy conditions on the simplest choice of space-time involving the Gromoll--Meyer sphere $\Sigma$: an $8$-dimensional static space-time whose space-like part is $\Sigma$. We end with Section \ref{sec:Conclusions}, which contains a summary and a discussion about future directions.

\section{Quaternions, instantons and spheres}
\label{sec:Quat_inst_and_spheres}
In this section, we summarise the quaternionic notation that will be used throughout the paper. We show how geometric quantities of physical interest can be recast in terms of quaternionic-valued objects by focussing on $SU(2)$ (multi-)instantons on $S^4$ and the vielbein of $S^3$, $S^4$. These are key ingredients of the Kaluza--Klein construction presented in \cite{Gherardini:2023uyx}, but their representation provided here is more natural and efficient for computations. We end this section by briefly discussing Milnor bundles.

\subsection{'t Hooft notation vs. quaternions}
\label{sec:'tHooft_notation_and_quaternions}
Although not frequently employed, instantons admit a very elegant description in terms of quaternions and quaternionic-valued forms. The representation that will be used throughout the paper is summarised by the following choice of basis:
\begin{align}
      \boldsymbol{e}_{c} =  ( I, - i \Vec{\tau}) \, , \quad \bar{ \boldsymbol{e}}_{c} = ( I, i \Vec{\tau}) \,  .
     \label{eq:Defn_e_0}
\end{align}
where $\Vec{\tau}$ are the Pauli matrices and $I$ is the $2 \times 2$ identity matrix. In this section only, we use bold symbols to denote quaternionic objects, to avoid any confusion.  With these definitions, we have the isomorphism with quaternions given by the map:
\begin{align}
    \boldsymbol{e}_{0} \xrightarrow{} \boldsymbol{1} \, , \quad
     \boldsymbol{e}_{1} \xrightarrow{} \boldsymbol{i} \, , \quad   \boldsymbol{e}_{2} \xrightarrow{} \boldsymbol{j} \, , \quad   \boldsymbol{e}_{3} \xrightarrow{} \boldsymbol{k}   \, ,
\end{align}
and the definition of $\bar{\boldsymbol{e}}_{c}$ is consistent with quaternionic conjugation. Let us denote quaternions with $\HH$, unit quaternions with $\HH^* = \{\boldsymbol{x}\in\HH:|\boldsymbol{x}|=1\}$ and imaginary quaternions with $\HH' = \{\boldsymbol{x}\in\HH:\mathrm{Im}(\boldsymbol{x}) = \boldsymbol{x} \iff \mathrm{Re}(\boldsymbol{x}) = 0 \}$; as usual, $\mathrm{Re}(\boldsymbol{x}) = \frac{1}{2} (\boldsymbol{x} + \bar{\boldsymbol{x}})$, $\mathrm{Im}(\boldsymbol{x}) = \frac{1}{2} (\boldsymbol{x} - \bar{\boldsymbol{x}})$ and $|\boldsymbol{x}|^2=\boldsymbol{x} \bar{\boldsymbol{x}}$. Then, the standard isomorphisms read:
\begin{align}
    \HH \simeq \mathbb{R}^4 \, , \quad  \HH^* \simeq S^3 \simeq SU(2) \, , \quad \HH'  \simeq \su(2) \, .
    \label{eq:HH_isomorphisms}
\end{align}
Accordingly, coordinates $x^{m}$ on $\RR^4$ can be organised into a quaternionic object as 
$ \boldsymbol{x} = x^{m} \boldsymbol{e}_m $. 
The exterior derivative is given by 
$\mathrm{d} \boldsymbol{x} = \dd x^{m} \boldsymbol{e}_m $, 
as expected.
Then, the expression for the usual $k=1$ instanton field strength in regular gauge is
\begin{align}
    \boldsymbol{F}={\lambda^2\over(\lambda^2+|\boldsymbol{x}-\boldsymbol{\xi}|^2)^2} \dd \boldsymbol{x} \wedge \dd\bar{\boldsymbol{x}} \, ,
    \label{eq:Quaternionic_F_k=1}
\end{align}
where $\boldsymbol{\xi}$ is a constant quaternion, containing the position moduli, and the wedge product is defined by antisymmetrisation of component 1-forms $\dd x^m$ and quaternionic multiplication. To show the equivalence of this expression to the usual one, it is sufficient to realise that $\boldsymbol{e}_{[m} \bar{\boldsymbol{e}}_{n]} = {1\over2} (\boldsymbol{e}_{m} \bar{\boldsymbol{e}}_{n} - \boldsymbol{e}_{n} \bar{\boldsymbol{e}}_{m})$ is selfdual, \ie, 
\begin{align}
\boldsymbol{e}_{[m} \bar{\boldsymbol{e}}_{n]}=\frac{1}{2} \epsilon_{m n p q} \boldsymbol{e}_{[p} \bar{\boldsymbol{e}}_{q]}\;.
\label{eq:selfduality}
\end{align}
 The components of the form in \eqref{eq:Quaternionic_F_k=1} read:
\begin{align}
     \boldsymbol{F}_{m n} = {2\lambda^2\over(\lambda^2+|\boldsymbol{x}-\boldsymbol{\xi}|^2)^2} \boldsymbol{e}_{[m} \bar{\boldsymbol{e}}_{n]}  \, ,
\end{align}
matching the standard expression for the $SU(2)$ 1-instanton, up to re-labelling of indices, as we shortly discuss.\footnote{Tensors of the form $\boldsymbol{e}_{[m} \bar{\boldsymbol{e}}_{n]}$, which might differ by permutations and minus signs, are often denoted as $\sigma_{m n}$ in the literature - see for instance \cite{Vandoren:2008xg}.} A key point is that $\boldsymbol{e}_{[m} \bar{\boldsymbol{e}}_{n]}$ is an imaginary quaternion for any $m, n$ and, as depicted in \eqref{eq:HH_isomorphisms}, imaginary quaternions form a representation of $\su(2)$ - the fundamental; this can be seen directly from \eqref{eq:Defn_e_0}. When extracting the components of the field strength in the basis $\{ \boldsymbol{e}_i \}$, $i=1,2,3$, one does not find the standard 't Hooft symbols. This is because of our choice in \eqref{eq:Defn_e_0}, where the real part was (naturally) labelled as the zeroth component. Instead, one finds the \textit{reversed} 't Hooft symbols, which read:
\begin{align}
& ^o\eta_{i m n} =\epsilon_{i m n 0} -\delta_{i m} \delta_{n 0} + \delta_{i n} \delta_{m 0} \, , \nonumber \\
& ^o\bar{\eta}_{i m n}=\epsilon_{i m n 0} + \delta_{i m} \delta_{n 0} - \delta_{i n} \delta_{m 0} \, ,
\label{eq:'tHooft_zero}
\end{align}
where $^o\eta_{i m n}$ is selfdual and $^o\bar{\eta}_{i m n}$ is anti-selfdual. The reverse 't Hooft symbols differ by the standard ones by moving the zeroth component to the fourth position (which also exchanges selfdual with anti-selfdual). Hence, the component expression of the field strength for a $k=1$ instanton in our conventions reads
\begin{align}
    F^i_{m n} = -  {2\lambda^2\over(\lambda^2+|\boldsymbol{x}-\boldsymbol{\xi}|^2)^2}  \left. ^o\eta^i_{m n} \right.  \, .
\end{align}

The associated gauge field, obeying $\boldsymbol{F}=d\boldsymbol{A}+\boldsymbol{A}\wedge \boldsymbol{A}$, is given by 
\begin{align}
\boldsymbol{A}={\Im( (\boldsymbol{x} - \boldsymbol{\xi})\dd \bar{\boldsymbol{x}})\over({\lambda^2 + |\boldsymbol{x} - \boldsymbol{\xi}|^2})}
 \;.
\label{eq:RegularA_k=1}
\end{align}
To show that the component expression also matches the classic BPST instanton of \cite{BELAVIN197585, tHooft:1976snw}, up to relabelling, one shall use the identity $(\boldsymbol{x} \bar{\boldsymbol{y}})_i =- \, ^o\eta_{i m n} x^{m} y^{n}$, where $(\cdot)_i$ indicates the $i^{\mathrm{th}}$ component of the quaternion, to find:
\begin{align}
    A^i_{m} = - {1\over\lambda^2+|\boldsymbol{x}-\boldsymbol{\xi}|^2}  \left. ^o\eta^i_{m n} (x^n - \xi^n) \right. \, .
\end{align}
On the other hand, to go from \eqref{eq:RegularA_k=1} to \eqref{eq:Quaternionic_F_k=1}, it is convenient to use the relations
\begin{align}
    \begin{aligned}
& -2 \operatorname{Re} \dd\boldsymbol{x} \wedge \operatorname{Im} \dd\boldsymbol{x}-\operatorname{Im} \dd\boldsymbol{x} \wedge \operatorname{Im} \dd\boldsymbol{x}=\dd\boldsymbol{x} \wedge \dd\bar{\boldsymbol{x} } \, , \\
& -4 \operatorname{Re} \dd\boldsymbol{x} \wedge \operatorname{Im}\dd \boldsymbol{x}+\dd\bar{\boldsymbol{x}} \wedge \dd\boldsymbol{x}=\dd\boldsymbol{x} \wedge \dd\bar{\boldsymbol{x}} \, .
\end{aligned}
\label{eq:Quat_identity_1}
\end{align}
We collect all of these, and other useful formulae for quaternionic computations, in Appendix \ref{sec:Appendix_B}.

\subsection{Background geometry\label{BGSection}}

All bundles we will consider are constructed with the round $S^4$ as base space and the round $S^3$ as fiber.
The radius of $S^3$ will always be 1. To encode the relative size of the spheres, we (sometimes) introduce a radius $r$ for $S^4$. Most calculations are performed for $r=1$, the results can then be scaled appropriately.

Let us briefly sketch how the ``background geometry'' $S^4\times S^3$ is dealt with in quaternionic language.

As mentioned, we view $S^3\simeq SU(2)$ as the space of unit quaternions $\{y\in\HH:|y|=1\}$ (note that we have dropped the bold notation). We will not bother to divide $S^3$ in coordinate patches. 
The vielbein can be seen as a 1-form taking values in $\HH'$, the imaginary quaternions; it reads
\begin{align}
\varepsilon=\dd y\bar y=-y\dd\bar y\;,
\end{align}
and $\dd s^2=\Re(\varepsilon\otimes\bar\varepsilon)$.
It fulfills the Maurer--Cartan equation $\dd\varepsilon-\varepsilon\wedge\varepsilon=0$.
The spin connection, also an imaginary $1$-form (\ie, an $\su(2)$-valued 1-form), fulfills the vanishing torsion condition
$\dd\varepsilon+\omega\wedge\varepsilon+\varepsilon\wedge\omega=0$. We thus have 
\begin{align}
\omega=-{1\over2}\varepsilon=-{1\over2}\dd y\bar y\;.
\end{align}
The curvature is $r=\dd\omega+\omega\wedge\omega=-{1\over4}\varepsilon\wedge\varepsilon$, with components
$r_{ij}{}^k=-{1\over2}\epsilon_{ij}{}^k$. Translating the index $k$ to an antisymmetric pair according to 
``$v^{ij}=-2\epsilon^{ij}{}_kv^k$'' gives $r_{ij}{}^{kl}=2\delta_{ij}^{kl}$, appropriate for a sphere with radius 1.

The isometry $SO(4)\simeq(SU(2)\times SU(2))/\ZZ_2$ of $S^3$ is realised as left and right action with unit quaternions:
$y\mapsto uy\bar v$. Notice that the choice of $\varepsilon$ above amounts to choosing the right-invariant Maurer--Cartan forms. We might as well have chosen the left-invariant ones $\epsilon'=\bar y\dd y=-\dd\bar y y$. The translation between them by conjugation with $y$ will be the source of explicit $y$-dependence in the Kaluza--Klein construction.

The $S^4$ is described in two patches, each excluding one pole of $S^4$. For each patch, we note that $\RR^4\simeq\HH$ and we use a coordinate $x\in\HH$, with the overlap $x'=x^{-1}$ between the patches. 
The metric for radius 1 is encoded in $\dd s^2=\Re(E\otimes\bar E)$ with the vielbein
\begin{align}
E={2\dd x\over1+|x|^2}\;.
\end{align}
The local $\so(4)\simeq\su(2)\oplus\su(2)$ acts by left and right multiplication by elements in $\HH'$.
The vanishing torsion condition then reads
$\dd E+\Omega_L\wedge E+E\wedge\Omega_R=0$, which is solved by
\begin{align}
\Omega_L&={\Im(x\dd\bar x)\over1+|x|^2}\;,\nn\\
\Omega_R&={\Im(\bar x\dd x)\over1+|x|^2}\;.
\end{align}
One finds left (selfdual) and right (anti-selfdual) curvatures
\begin{align}
R_L&={\dd x\wedge\dd\bar x\over (1+|x[^2)^2}={1\over4}E\wedge\bar E\;,\nn\\
R_R&={\dd \bar x\wedge\dd x\over (1+|x[^2)^2}={1\over4}\bar E\wedge E\;.\label{leftrightS4Riemann}
\end{align}
Their sum, translated from $\HH'\oplus\HH'$ to antisymmetric pairs of indices, is in flat indices
$R_{ab}{}^{cd}=2\delta_{ab}^{cd}$. The left and right spin connections are connections on the 1-instanton and 1-anti-instanton bundles on $S^4$ (see Section \ref{sec:KKModuli_vs_Instanton_moduli}).

In order to derive the action of the $SO(5)$ isometry on $x$, we start from the homogeneous coordinates of $\HH P^1=S^4$:
\begin{align}
Z=\left(\begin{matrix}z_1\\z_2\end{matrix}\right)\in\HH^2
\end{align}
on which the isometry group acts linearly, $Z\mapsto MZ$. $M$ fulfills $MM^\dagger=I=M^\dagger M$, and is a group element in $U(2,\HH)\simeq USp(4)\simeq Spin(5)$:
\begin{align}
M=\left(\begin{matrix}a&b\\c&d\end{matrix}\right)\;,\quad |a|^2+|b|^2=1=|c|^2+|d|^2\;,\quad a\bar c+b\bar d=0\;.
\label{so5matrix}
\end{align}
Note that the relations imply $|a|^2=|d|^2$, $|b|^2=|c|^2$.
$S^4=\HH P^1$ is obtained from the homogeneous coordinates $Z$ as $\HH^2/\HH=\HH^2/(SU(2)\times\RR_+)$, where the orbits are generated as $Z\mapsto Z\alpha$, $\alpha\in\HH$, which commutes with left multiplication by $M$.
In the patch where $z_2\neq0$, we can choose a representative 
\begin{align}
Z=\left(\begin{matrix}x\\1\end{matrix}\right)\;,
\end{align}
leading to a quaternionic M\"obius transformation
\begin{align}
\left(\begin{matrix}x\\1\end{matrix}\right)
\mapsto\left(\begin{matrix}ax+b\\cx+d\end{matrix}\right)
\approx\left(\begin{matrix}(ax+b)(cx+d)^{-1}\\1\end{matrix}\right)\;.
\end{align}
The linearly realised $SO(4)$ subgroup is described by diagonal matrices with $|a|=1=|d|$, and $x\mapsto ax\bar d$. While this transformation leaves the form of the round metric on $S^4$ invariant, the same is not true for the expressions of $k=1,2$ instantons. Hence, the $SO(5)$ isometries of the base act non-trivially on the moduli space of the instantons. This, in turn, implies that the moduli space of the instanton is not the moduli space of the Kaluza--Klein metric, since the action of the isometry group $SO(5)$ needs to be quotiented out. This point will be discussed in more detail in Section \ref{sec:KKModuli_vs_Instanton_moduli}, and it is a key observation in order to identify special points in the moduli space of instantons.

\subsection{Left and right bundles and exotic spheres}
$SU(2)$ instantons on $S^4$ are characterised by the instanton number\footnote{There may be sign differences due to conventions across the literature, due to \eg\ definition of dualisation. In our conventions, selfdual solutions according to eq. \eqref{eq:selfduality} have positive instanton number.}
\begin{align}
k=-{1\over4\pi^2}\int_{S^4}\Re(F\wedge F)\;.\label{eq:InstantonNumber}
\end{align}
Evaluating this integral, we need to use the two patches of $S^4$. Let the gauge transformation on the overlap be $g$, so that 
$A'=g\dd g^{-1}+gAg^{-1}$, $F'=gFg^{-1}$. The integral can then be written as 
\begin{align}
k=&-{1\over4\pi^2}\int_{S^3}\Re(A\wedge\dd A+{2\over3}A\wedge A\wedge A)\nn\\
&+{1\over4\pi^2}\int_{S^3}\Re(A'\wedge\dd A'+{2\over3}A'\wedge A'\wedge A')\;,
\label{kCS}
\end{align}
where the coboundary $S^3$ is contained in both patches. A standard choice is the unit sphere, $|x|=1$. Using the gauge transformation,
this turns into
\begin{align}
k={1\over12\pi^2}\int_{S^3}\Re(g^{-1}\dd g)^3\;.
\end{align}
This is minus the winding number of $g$ on $S^3$. Take \eg\ $g=x$ ($|x|=1$).
We can think of $e=g^{-1}\dd g$ as a vielbein on $S^3$ of radius 1 (see Section \ref{BGSection}).
Then, $\Re(g^{-1}\dd g)^3=-\epsilon_{ijk}\dd x^m\wedge\dd x^n\wedge\dd x^pe_m{}^ie_n{}^je_p{}^k$, and
$\star\Re(g^{-1}\dd g)^3=-6$. The integral becomes $k={1\over12\pi^2}\times(-6)\times2\pi^2=-1$. 

For a selfdual $F$, \eqref{eq:InstantonNumber} becomes $k={1\over8\pi^2}\int_{S^4}d^4x\sqrt gg^{mp}g^{np}F_{mn}{}^iF_{pq}{}^i
=\int_{S^4}d^4x\sqrt g\II$. We refer to $\II={1\over8\pi^2}g^{mp}g^{np}F_{mn}{}^iF_{pq}{}^i$ as the instanton scalar, and $\sqrt g\II$ as the instanton density.

When constructing $S^3$ bundles over $S^4$, there are two $SU(2)$'s present, acting on the unit quaternion $y$ parametrising $S^3$ by left and right multiplication. Both can be twisted on the overlap as above, leading to instantons for each $SU(2)$. 
These bundles are so called Milnor bundles \cite{10.2307/1969983}, with overlaps
\begin{align}
x'&=x^{-1}\;,\nn\\
y'&=e^{-m}ye^{n}\;,\label{Overlap}
\end{align}
where $y\in\HH$, $|y|=1$ parametrises $S^3$, $x\in\HH$ parametrises $\RR^4$, and $e={x\over|x|}$.
With this definition of the integer winding numbers, $m$ and $n$ coincide with the instanton numbers of the left and right $SU(2)$, respectively. Note that this labelling differs from the original one employed by Milnor, where the two integers ($h=-m$ and $l=n$) correspond to the powers of the quaternions, and there is no minus sign involved.

There are a priori two copies of $SO(4)\simeq (SU(2)\times SU(2))/\ZZ_2$, where the $SU(2)$'s act by left and right multiplication by unit quaternions on $x$ and $y$:
\begin{align}
x&\mapsto\alpha x\bar\beta\;,\nn\\
y&\mapsto\gamma y\bar\delta\;.
\end{align}
A selfdual 2-form with basis elements $\dd x\wedge \dd \bar x$ transforms only under $SU(2)_\alpha$:
$\dd x\wedge \dd \bar x\mapsto\alpha \dd x\wedge \dd \bar x\bar\alpha$, and anti-selfdual $\dd \bar x\wedge \dd x$ under $SU(2)_\beta$.
$m>0$ in eq. \eqref{Overlap}, and also $n>0$, corresponds to selfdual instantons, and $m,n<0$ to anti-selfdual instantons.
Conjugation of $x$ interchanges $SU(2)_\alpha\leftrightarrow SU(2)_\beta$, and conjugation of $y$ interchanges
$SU(2)_\gamma\leftrightarrow SU(2)_\delta$. 
Thus, from eq. \eqref{Overlap}, $x$-conjugation gives $(m,n)\mapsto(-m,-n)$, instantons are interchanged with anti-instantons,
while $y$-conjugation gives $(m,n)\mapsto(n,m)$. 

Milnor showed in \cite{10.2307/1969983}, via Morse theory, that when $-m + n = 1$, the total space of the  bundle is homeomorphic to the topological $7$-sphere; $-m + n = -1$ also guarantees the existence of a homeomorphism, by the same argument or just by realising that $m,n \xrightarrow{} -m,-n$ is an orientation reversing isomorphism of vector bundles (this is the $x$-conjugation mentioned above, see \cite{McEnroe2016MILNORSCO} for a detailed account). Moreover, when $(m+n)^2 \neq 1(\bmod 7)$, the total space cannot be diffeomorphic to the ordinary $S^7$ - which is obtained as $m=1$, $n=0$. Hence, when both conditions are met, we are in the presence of an exotic sphere.
The simplest such case is $m=2$, $n=1$; we will investigate its geometry in some detail.\footnote{These windings are opposite to those appearing in \cite{Gherardini:2023uyx}. This does not really make a difference, since the two choices are related simply by an $x$-conjugation, or change of orientation.}

The left- and right-twisted bundles with instanton numbers $(m,n)$ can be obtained from the (principal) $(S^3\times S^3)$-bundle with
\begin{align}
x'&=x^{-1}\;,\nn\\
y'&=e^{-m}y\;,\\
z'&=e^{-n}z\;\nn
\end{align}
by modding out $(y,z)\approx(y\bar\delta,z\bar\delta)$. Choose $(y,1)$ as a representative. Then,
$(y',z')=(e^{-m}y,e^{-n})\approx(e^{-m}ye^{n},1)$, and the Milnor bundle is obtained.

Let us start from a metric $ds^2=\Re(E\otimes\bar E+\varepsilon\otimes\bar\varepsilon+\varphi\otimes\bar\varphi)$, where $E$ is the quaternionic vielbein on $S^4$, and $\varepsilon$ and $\varphi$ are imaginary quaternionic vielbeins on the two $S^3$'s, with
\begin{align}
\varepsilon&=a(\dd y\bar y+A)\;,\nn\\
\varphi&=b(\dd z\bar z+B)\;.
\end{align}
$a$ and $b$ are the radii of the $S^3$'s, and $A$ and $B$ are $SU(2)$ connections (\ie, $A=\dd x^mA_m{}^i(x)e_i$ etc.) with instanton numbers $m$ and $n$. It is then straightforward to calculate the distance between orbits. It is encoded in the new vielbein $\tilde\varepsilon$ on $S^3$ parametrised by $y$:
\begin{align}
\tilde\varepsilon={ab\over\sqrt{a^2+b^2}}(\dd y\bar y+A-yB\bar y)\;.
\end{align}
Taking $a=b=\sqrt2$ gives the vielbein we will use in the following, with unit radius for $S^3$.

\section{The Kaluza--Klein metric and its curvature}
\label{sec:Curvature}
In this section, we present the Kaluza--Klein metric in the quaternionic notation just outlined, by making the connection with component expressions of \cite{Gherardini:2023uyx} explicit. Moreover, we comment on the role of some special diffeomorphisms of the total space, which will be considered in the next sections. Then, we perform the calculation of the Riemann tensor, Ricci curvature and Ricci scalar, finding an agreement with existing results in the literature.

\subsection{Metric and isometries}
Let the metric on the total space of the bundle be $ds^2=E^a\otimes E^a+\varepsilon^i\otimes\varepsilon^i$ ($a=0\ldots3$, $i=1\ldots3$), where we write 
$\varepsilon=\varepsilon^ie_i$ as a 1-form in $\HH'$, the imaginary quaternions:
\begin{align}
E^a&=\dd x^mE_m{}^a \, , \nn \\
\varepsilon&=\epsilonzero+A-yB\bar y\;.
\label{eq:Vielbein_quaternions}
\end{align}
Following the notation of Section \ref{BGSection}, we use $E_m{}^a(x)$ to denote the vielbein on an $\RR^4$ patch of $S^4$, and $\epsilonzero$ for the vielbein on the round $S^3$ ($\epsilonzero = \dd y\bar{y}$, $|y|=1$). 
$A$ and $B$ are connections for the left and right $SU(2)$ isometries on $S^3$.

Let us now briefly show the equivalence of this Ansatz with the one of \cite{Gherardini:2023uyx}. The components of the metric therein are given by:
\begin{align}
    \left(\begin{array}{cc}
g_{m n}(x) + 4\delta_{i j} \mathchorus{K}_{\,\, I}{}^i(y) \mathchorus{K}_{\,\, J}{}^j(y) A_m^I(x) A_n^J(x)   \,\,\,
&  4A_m^j(x) 
 + 4A_m^{\hat{i}}(x) \mathchorus{K}_{\,\,\hat{i}}{}^j(y)  \\
 4A_{n}^{i}(x) + 4A_{n}^{\hat{i}}(x) \mathchorus{K}_{\,\,\hat{i}} {}^i(y) & 4\delta_{i j}
\end{array}\right),
\label{eq:Expansion_metric_1}
\end{align}
where we have re-labelled some indices to make the notation consistent with the choices above, so that $I = ( i, \hat{i} ) = 1,\dots, 6$ refer to the two $su(2)$ components of the Lie algebra $so(4)$; accordingly, $A_{m}^{i}$ are the components of $A$ and $A_m^{\hat{i}}$ are components of $B$. The factors of $4$ are simply due to an unconventional choice of the generators of $SU(2)$ (see \cite{Gherardini:2023uyx} for more details), and the reader is referred to \cite{10.1063/1.525753} for a thorough derivation of the general Ansatz. It is worth noting that, upon the choice a of bi-invariant metric on the fibre, which identifies the right- and left-invariant vector fields with the Killing vectors $\xi_{I}{}^{\tau}$ ($\tau$ being the curved index on $S^3$, with coordinates $z^{\tau}$), then one can re-write the Ansatz as:
\begin{align}
    \mathrm{d} s^2=\left(E_{m}^{a} \mathrm{d} x^{m}\right)^2+\left(\varepsilon^i{}_\tau \mathrm{d} z^\tau- \varepsilon^i{}_{\tau} \, \xi_{I}{}^\tau A_{m}^{I} \mathrm{d} x^{m}\right)^2 \, .
    \label{eq:Classic_KK_ansatz}
\end{align}
This is a more common expression for the Kaluza--Klein Ansatz within the physics literature (see \cite{DUFF198490, Bailin:1987jd, Salam:1981xd}, for instance). Before discussing its isometries, let us quickly return to \eqref{eq:Expansion_metric_1} to expose its equivalence with \eqref{eq:Vielbein_quaternions}. In \cite{Gherardini:2023uyx}, the unit $S^3$ was embedded in $\RR^4$ as $\{ (X, Y, Z, W ) : X^2 + Y^2 + Z^2 + W^2 = 1 \}$, which yields $\mathchorus{K}_{\,\, i }{}^{j} = \delta_{i }{}^{j}$ and
\begin{align}
    \mathchorus{K}_{\,\,\hat{i}}{}^{i} =
\left(
\begin{array}{ccc} \vspace{0.1 cm}
 1-2 \left(W^2+X^2\right) & -2 (W Z+X Y) & 2 W Y-2 X Z \\ \vspace{0.1cm}
 2 (W Z-X Y) & 1-2 \left(W^2+Y^2\right) & -2 (W X+Y Z) \\
 -2 (W Y+X Z) & 2 W X-2 Y Z & 2 \left(X^2+Y^2\right)-1 
\end{array}
\right)_{\hat{i}}^{\,\,\,\, i } .
\label{eq:K_in_left_right_basis}
\end{align}
With the identification $y=(-W,X,Y,Z)_c \, e_c$, one finds that $(y B \bar{y})^{i} = -A_m^{\hat{i}}(x) \mathchorus{K}_{\,\,\hat{i}}{}^i \mathrm{d}x^m$, which proves the equivalence. Note the efficiency of the quaternionic notation, where the whole matrix \eqref{eq:K_in_left_right_basis} is encoded by the conjugation by $y$ in eq. \eqref{eq:Vielbein_quaternions}.

Let us now turn our attention to the diffeomorphisms of the total space of the bundles that we are considering. Some of them are particularly relevant within the Kaluza--Klein construction: they are the isometries of the base and the base-dependent isometries of the fibre. We start by discussing the former in the specific context of our investigation. 

Isometries of the base play a role in the ``inverse" construction that we are focussing on, where they determine one (or more) natural choice(s) of connection on the bundle. Concretely, the round metric on the base manifold $S^4$ is invariant under $SO(5)$ transformations, which were reviewed in Section \ref{BGSection}. These transformations, however, do not necessarily leave the gauge field unchanged. Hence, all of those gauge field configurations that are related by $SO(5)$ transformations should be identified for our purposes, since plugging them into the Kaluza--Klein Ansatz just produces diffeomorphic metrics on the total space. As discussed in Section \ref{sec:KKModuli_vs_Instanton_moduli}, the case of $k=1$ instantons is special, in that there exist a choice of moduli which is fixed point of the $SO(5)$ action. This is therefore a reasonable choice for the connection on the bundle, which is always made in all the constructions of the round metric on $S^7$ treated as a quaternionic Hopf fibration. For the $k=2$ case, things are a bit more subtle, since there is no fixed point. This is also discussed in the next Section.

Isometries of the fibre, on the other hand, have been discussed thoroughly in the literature, and we just quickly review them here. To do that, it is convenient to consider the metric \eqref{eq:Classic_KK_ansatz}. Then, non-Abelian gauge transformations arise by considering the effect on the components $\bar{g}_{m \tau}$ of the infinitesimal isometry of the fibre metric $g_{\tau \omega }$, with $x$-dependent parameters:
\begin{align}
    z^{\tau} \rightarrow z^{\tau}+\xi_{I}^{\tau}(y) \theta^{I}(x) \, .
\end{align}
One finds that:
\begin{align}
    A_{m}^{I} \rightarrow A^{'}{}_{m}^{I}=A_{m}^{I}+\partial_{m} \theta^{I}+C_{IJK} \theta^{J} A_m^{K} \, 
\end{align}
where $C_{IJK}$ are the structure constants of the algebra associated to the isometries of the fibre, \ie, $\so(4)\simeq\su(2)\oplus\su(2)$ for us (since the fibre is $S^3$). Hence, base-dependent isometries of the fibre effectively implement gauge transformations on the connection of the bundle, as one would expect from the Kaluza--Klein Ansatz.

\subsection{Bundle vielbein, connection and curvature\label{BundleGeometry}}
We want to find the spin connections, and then the curvature, associated with $\eqref{eq:Vielbein_quaternions}$. 
Let us divide the $\so(7)$ spin connection in three parts, depending on the index structure, schematically
\begin{align}
\left(\begin{matrix}\Omega&-\nu^\transpose\\ \nu&\omega\end{matrix}\right)\;.
\end{align}
Let us begin with $\omega$, the $\so(3)$ spin connection on $S^3$. 
It is convenient to represent it as a 1-form in $\HH'$.
It is
\begin{align}
\omega=\omegazero+{1\over2}(A+yB\bar y)\;,
\end{align}
where $\omegazero$ is the connection on the round $S^3$ defined in Section \ref{BGSection}.
Note the different relative sign of $A$ and $B$ compared to $\varepsilon$.
It is then straightforward to verify that
\begin{align}
d\varepsilon+\omega \wedge \varepsilon + \varepsilon \wedge \omega
=F-yG\bar{y}\equiv\sF\;,
\end{align}
with $F=dA+A\wedge A$, $G=dB+B\wedge B$. This comes from an interplay between terms with different signs:
\begin{align}
d\varepsilon+\omega\wedge\varepsilon+\varepsilon\wedge\omega
&=d\epsilonzero+\omegazero\wedge\epsilonzero+\epsilonzero\wedge\omegazero\nn\\
&\qquad+dA-ydB\bar y-\dd y\wedge By+yB\wedge \dd \bar y\nn\\
&\qquad+\omegazero\wedge(A-yB\bar y)+(A-yB\bar y)\wedge\omegazero\nn\\
&\qquad+{1\over2}(A+yB\bar y)\wedge\epsilonzero+{1\over2}\epsilonzero\wedge(A+yB\bar y)\\
&\qquad+{1\over2}(A+yB\bar y)\wedge(A-yB\bar y)+{1\over2}(A-yB\bar y)\wedge(A+yB\bar y)\nn\\
&=dA+A\wedge A-y(dB+B\wedge B)\bar y\;,\nn
\end{align}
where we have used $\dd y=\epsilonzero y$, $\dd \bar y=-\bar y\epsilonzero$ on the second line and
$\omegazero=-{1\over2}\epsilonzero$ on the third line.
Similar statements relate the spin connection on $S^3$ to the gauge connections. Let $X(x),Y(x)\in\HH'$, and
let $Z=X-yY\bar y$, $\tilde Z=X+yY\bar y$. Then,
\begin{align}
D^{(\omega)}Z&=dZ+[\omega,Z] \label{DZeq}\\
&=D^{(A)}X-yD^{(B)}Y\bar y-{1\over2}[\varepsilon,X+yY\bar y]\\
&\equiv DZ-{1\over2}[\varepsilon,\tilde Z]\nn
\end{align}
 by a similar calculation.

The remaining parts of the spin connection are
\begin{align}
\nu_{ia}&={1\over2}\imath_a\sF^i\;,\nn\\
\Omega_{ab}&=\Omegazero_{ab}-{1\over2}\sF_{ab}{}^i\varepsilon^i\;,
\end{align}
where $dE^a+\Omegazero{}^a{}_b\wedge E^b=0$.

The corresponding three parts of the Riemann tensor, decomposed as
\begin{align}
\left(\begin{matrix}R&-\varrho^\transpose\\ \varrho&r\end{matrix}\right)\;,
\end{align}
are
\begin{align}
R&=d\Omega+\Omega\wedge\Omega-\nu^\transpose\wedge\nu\;,\nn\\
\varrho&=d\nu+\nu\wedge\Omega+\omega\wedge\nu=D^{(\Omega,\omega)}\nu\;,\\
r&=d\omega+\omega\wedge\omega-\nu\wedge\nu^\transpose\;.\nn
\end{align}

In the resulting expressions, it is good to keep all $\dd y$'s expressed by $\varepsilon$, in order to be able to read off the flat components of the Riemann tensor.  A good check is that the components obtained this way are gauge covariant.
\begin{align}
r&=-{1\over4}\varepsilon\wedge\varepsilon+{1\over2}\tilde\sF+{1\over4}\imath_a\sF\wedge\imath_a\sF\;.
\end{align}
where $\tilde\sF=F+yG\bar y$
(still expressed as a 2-form in $\HH'$). 

Expressing also $\varrho$ as a 2-form $\varrho_a$ in $\HH'$,
\begin{align}
\varrho_a={1\over2}(d\imath_a\sF+\Omega_{ab}\imath_b\sF+\omega\wedge\imath_a\sF+\imath_a\sF\wedge\omega)\;,
\end{align}
we can use eq. \eqref{DZeq} to get the result
\begin{align}
\varrho_a={1\over2}(D^{(\Omegazero,A)}\imath_aF-yD^{(\Omegazero,B)}\imath_aG\bar y)
-{1\over4}\sF_{ab}{}^j\varepsilon^j\wedge\imath_b\sF
-{1\over4}\varepsilon\wedge\imath_a\tilde\sF-{1\over4}\imath_a\tilde\sF\wedge\varepsilon\;.
\end{align} 
The final result is checked for the window (\yng(2,2)) symmetry, and is, in flat components:
\begin{align}
R_{ab,cd}&=\Rzero_{ab,cd}-{1\over2}\sF_{ab}{}^i\sF_{cd}{}^i+{1\over2}\sF_{a[c}{}^i\sF_{d]b}{}^i\;,\nn\\
R_{ab,ci}&=-{1\over2}D_c\sF_{ab}{}^i\;,\nn\\
R_{ab,ij}&=-\epsilon_{ijk}\tilde\sF_{ab}{}^k-{1\over2}\sF_{[a}{}^{ci}\sF_{b]c}{}^j\;,\nn\\
R_{ai,bj}&=-{1\over2}\epsilon_{ijk}\tilde\sF_{ab}{}^k+{1\over4}\sF_{a}{}^{cj}\sF_{bc}{}^i\;,
\label{RiemannTensor}\\
R_{ai,jk}&=0\;,\nn\\
R_{ij,kl}&=2\delta_{ij}^{kl}\;.\nn
\end{align}
The covariant derivative is with $\Omegazero$, $A$ and $B$ (and thus does not feel the $y$'s in $\sF$).
(We have reverted to the notation $R$ for all components.)

The Ricci tensor obtained from this Riemann tensor is
\begin{align}
R_{ab}&=\Rzero_{ab}-{1\over2}\sF_a{}^{ci}\sF_{bc}{}^i\;,\nn\\
R_{ai}&={1\over2}D^b\sF_{ab}{}^i=0\;, \label{eq:Ricci_general_components}
\\
R_{ij}&=2\delta_{ij}+{1\over4}\sF^{abi}\sF_{ab}{}^j\;.\nn
\end{align}

We will always keep the radius of $S^3$ to $1$. The relative size of $S^4$ and $S^3$ is encoded in the radius of $S^4$. 
Geometrical quantities are obtained by scaling the radius 1 results to radius $r$. 
Then, it is clear that \eg\ the part $R_{ab}$ of the Ricci tensor as well as $\sF_{ab}{}^i$, both with flat indices, scale as $r^{-2}$. 

We can check that the 1-instanton (see Section \ref{k1SubSec}) reproduces the round and squashed $S^7$. Let the $S^4$ have the round metric with radius $r$. Then, $\Rzero_{ab}={3\over r^2}\delta_{ab}$. Also, let $\sF=F$. A 1-instanton of unit size has 
\begin{align}
F={\dd x\wedge \dd \bar x\over(1+|x|^2)^2}={1\over4r^2}E\wedge\bar E\;,
\end{align}
so that $F_{ab}{}^i=-{1\over2r^2}\Re(e_a\bar e_be_i)$, which gives
$F_a{}^{ci}F_{bc}{}^i={3\over4r^4}\delta_{ab}$, $F^{abi}F_{ab}{}^j={1\over r^4}\delta_{ij}$.
The non-vanishing parts of the Ricci tensor are
\begin{align}
R_{ab}&=({3\over r^2}-{3\over8r^4})\delta_{ab}\;,\nn\\
R_{ij}&=(2+{1\over4r^4})\delta_{ij}\;.
\end{align}
The metric is Einstein for $r={1\over2}$ and $r={\sqrt5\over2}$, with $R_{AB}=k\delta_{AB}$, $k=6$ and ${54\over25}$ respectively. The former case is the round $S^7$ with radius 1, and the latter the squashed $S^7$. It can be checked that, in the case $r={1\over2}$, the expressions in eq. \eqref{RiemannTensor} give $R_{AB,CD}=2\delta_{AB}^{CD}$, where $A=(a,i)$, so the sectional curvature is identically 1.

\section{Instanton moduli and Kaluza--Klein moduli}
\label{sec:KKModuli_vs_Instanton_moduli}

In this section, we comment on the symmetries and moduli spaces of the various geometric quantities that appear in the metric Ansatz just presented.
In what follows, we only consider bundles over a round $S^4$, which has isometry $SO(5)$.
This isometry will typically be broken by the presence of gauge connections $A$ and $B$. Additionally, the isometry of the round $S^3$ may be broken, partially or entirely.
Instanton solutions are parametrised by locus and size moduli (and relative $SU(2)$ orientation moduli for $k>1$, which we do not consider); these, however, do not coincide with the moduli of the space of geometric solutions.
As we already mentioned, if an instanton solution breaks part of $SO(5)$, the corresponding generators will transform the solution to other solutions. Since the ``geometric'' or ``Kaluza--Klein'' moduli should be counted modulo diffeomorphisms, the action of $SO(5)$ should be divided out, yielding a parameter space which is much smaller than the instanton moduli space. All of this is described in details below, together with the discussion of special choices in the moduli space.

Finally, in addition to instanton moduli, we also introduce a geometric modulus in the form of the radius of the base $S^4$. 

\subsection{Instanton solutions and moduli}

The $k$-instanton moduli space 
is the space of selfdual ($k>0$) or anti-selfdual ($k<0$) solutions with instanton number $k$. 
Note that the $\RR^4$ patches of $S^4$ are conformal to flat $\RR^4$. Dualisation of forms of degree ${d\over2}$ in $d$ dimensions only depends on the conformal class of the metric, so selfduality is the same on the round $S^4$ as on $\RR^4$. 
For instanton number $k>0$ the moduli space has dimension $8k-3$. It can be parametrised by $k$ loci, or ``centra'', $k$ (real) sizes and $k-1$ relative $SU(2)$ orientations, in total $4k+k+3(k-1)=8k-3$. Unlike instantons on $\RR^4$, where the moduli space has dimension $8k$, the overall orientation is a gauge parameter. The orientations may be combined with the sizes in quaternionic parameters, whose modulus is the size and whose ``phase'' is the orientation.

The most general method for finding instanton solutions (in any gauge) is the ADHM construction \cite{ATIYAH1978185}.
A somewhat simpler method, which does not capture the orientation moduli, is the method of harmonic functions \cite{PhysRevD.15.1642}.
We will not consider orientation moduli, the presence of which alters solutions significantly, so this method is in principle sufficient. 
It however has the drawback that connections and field strengths are given in ``singular gauge''.
Mathematically speaking, a singular gauge is not good, specifically it involves singularities (for the gauge connection and field strength) in each patch. Roughly speaking, in our previous terminology, the expression for $F'$ is used in the patch containing $x=0$. If one calculates the instanton number as in eq. \eqref{kCS}, the Chern--Simons integral can instead be localised close to the singularities.
In order to arrive at a ``regular gauge'', a ``singular gauge transformation'' has to be applied.
Even if the singular gauge is mathematically unsound, the expressions involved turn out to be somewhat simpler than the regular ones. The actual (regular) field strength can then be encoded in a singular one, together with the transformation that removes the singularity. For more details on the latter, see the following Subsections.

The construction from a harmonic function is straightforward. Let $\phi(x)$ be a harmonic function on 
$\RR^4\backslash\{a_1,\ldots,a_k\}$ with flat metric. Then a connection
\begin{align}
A=-{1\over2}\*_m\log\phi\,\Im(\bar e_m\dd x)
\end{align}
has a selfdual field strength
\begin{align}
F=-{1\over8}\bar e_m\dd x\wedge\dd\bar x e_n(\*_m\*_n\log\phi+\*_m\log\phi\,\*_n\log\phi)\;.
\end{align}
When the calculation is performed using quaternions, the crucial identity is (with $\*=e_m\*_m$)
$\*\bar\*\log\phi+(\*\log\phi)(\bar\*\log\phi)=0$. 
For a $k$-instanton, the harmonic function can be taken as
\begin{align}
\phi=1+\sum_{i=1}^k{\lambda_i^2\over|x-a_i|^2}\;,
\end{align}
where $\lambda_i$ are size moduli and $a_i$ location moduli (all different). This captures $5k$ of the $8k-3$ moduli on $S^4$.
The solutions are singular at $x=a_i$.

\subsection{$k=1$\label{k1SubSec}}

From the harmonic function $\phi=1+{\lambda^2\over|x-a|^2}$, we obtain the connection
\begin{align}
A={\lambda^2\Im(\bar x_a\dd x)\over|x_a|^2(\lambda^2+|x_a|^2)}\;,
\end{align} 
with $x_a=x-a$, and the field strength
\begin{align}
F={\lambda^2\bar x_a\dd x\wedge\dd\bar x x_a\over|x_a|^2(\lambda^2+|x_a|^2)^2}\;.
\end{align}
Clearly, the singularity at $x=a$ is an angular discontinuity in $F$ (but $A$, and hence the bundle metric, has a stronger singularity), which can be removed by a ``singular gauge transformation'' with parameter $g={x_a\over|x_a|}$. The regular connection and field strength are
\begin{align}
A'&=g\dd g^{-1}+gAg^{-1}={\Im(x_a\dd\bar x)\over\lambda^2+|x_a|^2}\;,\nn\\
F'&=gFg^{-1}={\lambda^2\dd x\wedge\dd\bar x\over(\lambda^2+|x_a|^2)^2}\; ,\label{k1solution}
\end{align}
as presented in Section \ref{sec:'tHooft_notation_and_quaternions}.

How does the isometry $SO(5)$ act on the moduli of instantons? Consider the field strength $F'$ as above,
with $\lambda\in\RR$ size modulus and $\xi\in\HH$ location moduli.
Under $SO(5)$ as in eq. \eqref{so5matrix},
\begin{align}
x&\mapsto(ax+b)(cx+d)^{-1}\;,\label{MobiusEq}\\
\dd x&\mapsto|cx+d|^{-2}(a-b\bar x)\dd x(cx+d)^{-1}\;,\nn
\end{align}
where the second transformation is obtained after a short calculation using the conditions on the matrix $M$.
Thus,
\begin{align}
\dd x\wedge \dd \bar x\mapsto|cx+d|^{-6}(a-b\bar x)\dd x\wedge \dd \bar x\overline{(a-b\bar x)}
=|cx+d|^{-4}u\dd x\wedge \dd \bar x\bar u\;,\label{dxdxtransformation}
\end{align}
where $u={a-b\bar x\over|cx+d|}$, which is a unit quaternion. We work with solutions modulo gauge transformations, so $u$ can be discarded.
We also need the transformation of the function in front in eq. \eqref{k1solution}, which becomes
\begin{align}
{\lambda^2\over(\lambda^2+|x-\xi|^2)^2}
\mapsto|cx+d|^4{\lambda^2\over (\lambda^2|cx+d|^2+|(a-\xi c)x+b-\xi d|^2)^2}
\end{align}
Rewriting this as $|cx+d|^4{\lambda'^2\over(\lambda'^2+|x-\xi'|^2)^2}$ involves one non-trivial check, that the same result for $\lambda'$ is obtained in the denominator and in the overall factor, so that one stays in the same class of $2$-forms, eq. \eqref{k1solution}. The factors of $|cx+d|$ are cancelled against those in eq. \eqref{dxdxtransformation}. The result is
\begin{align}
\xi\mapsto\xi'&=-{(a-\xi c)^{-1}(b-\xi d)+{\lambda^2\bar cd\over|a-\xi c|^2}\over1+{\lambda^2|c|^2\over|a-\xi c|^2}}\;,\nn\\
\lambda\mapsto\lambda'&={\lambda\over|a-\xi c|^2+\lambda^2|c|^2}\;.
\end{align}

Both $\xi'$ and $\lambda'$ in general depend on both $\xi$ and $\lambda$.  
The size modulus is not a scalar, and the location moduli do not transform with a simple M\"obius transformation, but one modified by $\lambda$. 
Transformations with $b=c=0$ act as expected, $\xi\mapsto\bar a\xi d$, $\lambda\mapsto\lambda$.
An instanton centered at $x=0$ transforms to
\begin{align}
\xi'&=-{\bar a b+\lambda^2\bar c d\over |a|^2+\lambda^2|c|^2}\;,\\
\lambda'&={\lambda\over  |a|^2+\lambda^2|c|^2}\;.\nn
\end{align}
A size $1$ instanton at $\xi=0$ is invariant (it is like a ``constant function'', being completely delocalised).
For any size modulus, one may always use an isometry to move a 1-instanton to be centered at $x=0$.

These considerations were based on the transformation of the field strength. 
It is quite instructive to elaborate on the transformation of $\dd x$ by itself.
A little calculation yields the transformation property
\begin{align}
{\bar x \dd x\over1+|x|^2}\mapsto(cx+d){\bar x \dd x\over1+|x|^2}(cx+d)^{-1}+(cx+d)\dd(cx+d)^{-1}\;.
\end{align}
This explains more or less directly the appearance of a gauge transformation of the connection (the one discarded above).

It is informative to examine the transformations under an ``inversion'', with the matrix $A$ having $a=d=0$. It sends the origin to infinity (so one needs to use the other patch, with coordinate $y=x^{-1}$).
We can view such a transformation as the limit of an element
\begin{align}
A_\beta={1\over\sqrt{1+|\beta|^2}}\left(\begin{matrix}1&\beta\\\bar\beta&-1\end{matrix}\right)
\end{align}
as $\beta\rightarrow\infty$. Let us take $\beta\in\RR$. Then, according to eq. \eqref{MobiusEq}, $x\mapsto x^{-1}$ (which is the coordinate transformation to the other patch).
If we let $A_\beta$ act on the moduli parameters $\xi=0$ and $\lambda$, however, the result is
\begin{align}
\xi'&={(\lambda^2-1)\beta\over1+\lambda^2|\beta|^2}\;,\nn\\
\lambda'&={\lambda(1+|\beta|^2)\over1+\lambda^2|\beta|^2}\;,
\end{align}
and the limit $\beta\rightarrow\infty$ is well defined. It agrees with the field strength \eqref{k1solution} with $\xi=0$ transformed to the other patch with the appropriate gauge transformation reflecting $k=1$:
With $x'=x^{-1}$, $\dd x=-x'^{-1}\dd x'x'^{-1}$, and
\begin{align}
F={\lambda^2\over(\lambda^2+|x'|^{-2})^2}x'^{-1}\dd x'x'^{-1}\wedge \bar x'^{-1}\dd \bar x'\bar x'^{-1}
={\lambda^{-2}\over(\lambda^{-2}+|x'|^2)^2}{\bar x'\over|x'|}\dd x'\wedge \dd \bar x'{x'\over|x'|}\;.
\end{align}
which again  has $\xi=0$ but size $\lambda'={1\over\lambda}$. An instanton centered at infinity is also centered at $0$, but with the inverse size.

The instanton scalar (here calculated when the center is at $x=0$) is 
\begin{align}
\II={1\over 8\pi^2}{(1+|x|^2)^4\over16}{4\lambda^4\times12\over(\lambda^2+|x|^2)^4}
={3\lambda^4\over8\pi^2}\Bigl({1+|x|^2\over\lambda^2+|x|^2}\Bigr)^4 \, ,
\label{eq:Inst_dens_k=1}
\end{align}
where the middle factor is $1/\sqrt g$ and the number $12$ comes from 
$\Re(e_{[a}\bar e_{b]}e^i)\Re(e_{[a}\bar e_{b]}e^i)=12$, see below.
The integral is of course $\int_{S^4}d^4x\sqrt g\II=1$.
For $\lambda=1$, $\II$ is constant over $S^4$.

In conclusion, we can always choose the locus to $0$. Then, the Kaluza--Klein moduli space only contains the size parameter $\lambda$, and only $\lambda\leq1$ (or $\lambda\geq1$). 
In the construction of the exotic $S^7$, it will be taken to $\lambda=1$ when centered at $x=0$, which is the only solution that does not break $SO(5)$.

\subsection{$k=2$}
\label{k2SubSec}

Let $x_a=x-a$, $x_b=x-b$. The singular gauge connection for a $k=2$ instanton, obtained from the harmonic function
$f=1+{\lambda_a^2\over|x_a|^2}+{\lambda_b^2\over |x_b|^2}$, is
\begin{align}
A={1\over1+{\lambda_a^2\over|x_a|^2}+{\lambda_b^2\over |x_b|^2}}\left(
{\lambda_a^2\Im(\bar x_a\dd x)\over|x_a|^4}+{\lambda_b^2\Im(\bar x_b\dd x)\over|x_b|^4}\right) \;.
\label{SingularA}
\end{align}
The field strength $F=dA+A\wedge A$ is easiest calculated in singular gauge, and then transformed to the regular one.
Given the form of eq. \eqref{SingularA}, it is clear that it will involve factors $\Re\omega$ and $\Im\omega$, where
$\omega=\bar x\dd x$ (with $x$ replaced by $x_a$ or $x_b$). One then uses identities like
\begin{align}
\Re\omega\wedge\Im\omega&=-{1\over4}\omega\wedge\bar\omega+{1\over4}\bar\omega\wedge\omega\;,\nn\\
\Im\omega\wedge\Im\omega&=-{1\over2}\omega\wedge\bar\omega-{1\over2}\bar\omega\wedge\omega\; ,
\end{align}
which are rearrangements of \eqref{eq:Quat_identity_1}, to arrive at the result
\begin{align}
F&={1\over(1+{\lambda_a^2\over|x_a|^2}+{\lambda_b^2\over|x_b|^2})^2}	
\left[{\lambda_a^2\over|x_a|^6}(1+{\lambda_b^2\over|x_b|^2})\bar x_a\dd x\wedge \dd \bar xx_a\right.\nn\\
&\qquad\left.+{\lambda_b^2\over|x_b|^6}(1+{\lambda_a^2\over|x_a|^2})\bar x_b\dd x\wedge \dd \bar xx_b
-{\lambda_a^2\lambda_b^2\over|x_a|^4|x_b|^4}(\bar x_a\dd x\wedge \dd \bar xx_b+\bar x_b\dd x\wedge \dd \bar xx_a)\right]\;,\\
&={1\over(\lambda_b^2|x_a|^2+\lambda_a^2|x_b|^2+|x_a|^2|x_b|^2)^2}
\left[\lambda_a^2|x_b|^2(\lambda_b^2+|x_b|^2){\bar x_a\dd x\wedge \dd \bar xx_a\over|x_a|^2}\right.\nn\\
&\qquad\biggl.+\lambda_b^2|x_a|^2(\lambda_a^2+|x_a|^2){\bar x_b\dd x\wedge \dd \bar xx_b\over|x_b|^2}
-\lambda_a^2\lambda_b^2(\bar x_a\dd x\wedge \dd \bar xx_b+\bar x_b\dd x\wedge \dd \bar xx_a)\biggr]\;.\nn
\end{align}
This explicitly displays selfduality with respect to a metric conformal to the flat metric on $\RR^4$, since all terms contain the selfdual $\dd x\wedge \dd \bar x$. It is clear that the singularities of $F$ at $x=a$ and $x=b$ are angular discontinuities. It can be checked
that $F'=gFg^{-1}$ is regular.
Terms in the scalar curvature and Ricci tensor are conveniently calculated in the singular gauge. Even though this corresponds to using coordinates with coordinate singularities at $x=a$ and $x=b$, the terms appearing in $R_{ab}$ are not affected by the gauge/coordinate transformation. The expressions for $R_{ij}$ need to be transformed to regular gauge.

The angular discontinuities in $F$ may be removed by the following ``singular gauge transformation'' \cite{Giambiagi:1977yg}.
Let $z={x_a\over|x_a|^2}-{x_b\over|x_b|^2}=\bar x_a^{-1}-\bar x_b^{-1}$. 
Then, 
\begin{align}
dz&=-\bar x_a^{-1}\dd \bar x\bar x_a^{-1}+\bar x_b^{-1}\dd \bar x\bar x_b^{-1}
=-{x_a\dd \bar xx_a\over|x_a|^4}+{x_b\dd \bar xx_b\over|x_b|^4}\;,\nn\\
|z|^2&={|a-b|^2\over|x_a|^2|x_b|^2}\;.
\end{align}
We want to make a gauge transformation with $g={z\over|z|}$. Then,
$A'=gdg^{-1}+gAg^{-1}$. We have
\begin{align}
gdg^{-1}&={\Im(z\dd \bar z)\over|z]^2}
=-{|x_b|^2\over |a-b|^2}{\Im(\dd x\bar x_a)\over|x_a|^2}-{|x_a|^2\over |a-b|^2}{\Im(\dd x\bar x_b)\over|x_b|^2}\nn\\
&+{1\over|a-b|^2}{\Im(x_a\bar x_b\dd x\bar x_b)\over|x_b|^2}
+{1\over|a-b|^2}{\Im(x_b\bar x_a\dd x\bar x_a)\over|x_a|^2}\;.\label{gdginveq}
\end{align}
Note that the divergence in the first term behaves as $-{\Im(\dd x\bar x_a)\over|x_a|^2}$ when $x\rightarrow a$, and similarly for the second term around $x=b$. Note also that the third and fourth terms are finite but not continuous at
$x=b$ and $x=a$, respectively. 
The other term in $A'$ is
\begin{align}
&gAg^{-1}=|z|^{-2}zA\bar z	\nn\\
&\quad={|x_a|^2|x_b|^2\over|a-b|^2}{1\over1+{\lambda_a^2\over|x_a|^2}+{\lambda_b^2\over |x_b|^2}}
   (\bar x_a^{-1}-\bar x_b^{-1})
   \left({\lambda_a^2\Im(\bar x_a\dd x)\over|x_a|^4}+{\lambda_b^2\Im(\bar x_b\dd x)\over|x_b|^4}\right)
   (x_a^{-1}-x_b^{-1})\nn\\
&\quad={|x_a|^2|x_b|^2\over|a-b|^2}{1\over1+{\lambda_a^2\over|x_a|^2}+{\lambda_b^2\over |x_b|^2}}\\
&\qquad\times\left[\lambda_a^2\left(
{\Im(\dd x\bar x_a)\over|x_a|^6}-{\Im(\dd x\bar x_b)\over|x_a|^4|x_b|^2}
+{\Im(x_b\bar x_a\dd x\bar x_b)\over|x_a|^4|x_b|^4}-{\Im(x_b\bar x_a\dd x\bar x_a)\over|x_a|^6|x_b|^2}\right)\right.\nn\\
&\left.\qquad\quad+\lambda_b^2\left({\Im(\dd x\bar x_b)\over|x_b|^6}-{\Im(\dd x\bar x_a)\over|x_a|^2|x_b|^4}
+{\Im(x_a\bar x_b\dd x\bar x_a)\over|x_a|^4|x_b|^4}-{\Im(x_a\bar x_b\dd x\bar x_b)\over|x_a|^2|x_b|^6}
\right)\right]\;.\nn
\end{align}
Number the terms (1)-(8) according to the position in the last parenthesis.
Terms (2), (3), (6) and (7) are regular at $x=a$ and $x=b$. The terms (1) and (4) are singular at $x=a$ and (5) and (8) at $x=b$.
At $x\approx a$, the behaviour of the singular terms (1) and (4) is
\begin{align}
(gAg^{-1})_{(1)+(4)}\approx
{\Im(\dd x\bar x_a)\over|x_a|^2}-{\Im((a-b)\bar x_a\dd x\bar x_a)\over|a-b|^2|x_a|^2}\;,
\end{align}
which cancels the behaviour of the first and fourth terms in $gdg^{-1}$, eq. \eqref{gdginveq}. In the same way, terms 
(5) and (8) cancel the singular behaviour of the second and third terms in $gdg^{-1}$.

The result is regular. It could of course be rewritten in a manifestly regular way, but we have no need for that expression.
If we examine the behaviour of $A'$ as $|x|\rightarrow\infty$, we find that the leading term comes entirely from 
$gdg^{-1}$ and is
\begin{align}
A'={\Im(x\dd \bar x)\over|x|^2}+{\Im(x(\bar a-\bar b)x\dd \bar x(a-b)\bar x)\over|a-b|^2|x|^4}+O(|x|^{-2})\;.
\end{align}

If we choose a frame where $a-b$ is real, this leading term equals $hdh^{-1}$, where $h={x^2\over|x|^2}$,
displaying the correct winding.

Extending the calculation of the $SO(5)$ transformations of moduli for $k=2$ seems complicated.
There will certainly be no $SO(5)$ fixed points in the $k=2$ moduli space. However, one will clearly always be able to transform the centra to (\eg) $\pm\xi$, $\xi\in\RR$ (or some similar desired relation if the sizes are different), so that the remaining parameters are two sizes and one distance (again, disregarding internal orientation). 

We will restrict our attention to equal size parameters. What does this mean, given the lesson from $k=1$ that size parameters are not scalar? We short-circuit this question by defining the class of equal-size 2-instantons as the solutions that are obtained from those with centra $\pm a$ and equal size $\lambda$ by an $SO(5)$ transformation. Then we will have no need for the explicit form of the other solutions in the orbits under $SO(5)$.
The only transformation still needed to divide out is the inversion.

\begin{figure}
\begin{center}
\includegraphics[scale=.5]{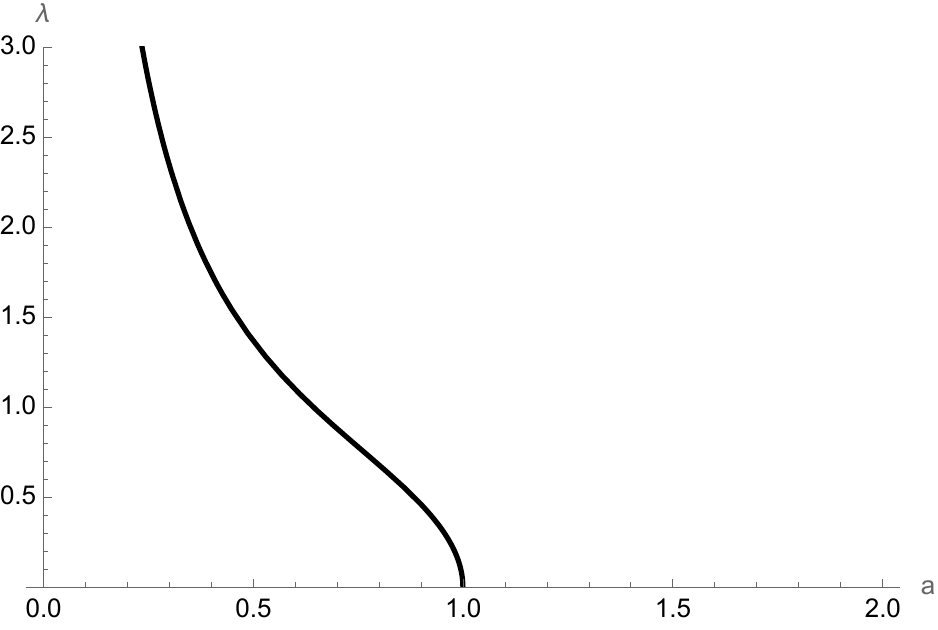}
\end{center}
\caption{The subspace of the geometric moduli space for equal size instantons. The curve separates the region of solutions with $\II(0)\geq\II(\infty)$ (left) from those with $\II(0)\leq\II(\infty)$ (right).}
\label{k2Figure}
\end{figure}

The field strength in a singular gauge is
\begin{align}
F&={\lambda^2\over(|x_+|^2|x_-|^2+\lambda^2|x_+|^2+\lambda^2|x_-|^2)^2}\nn\\
&\times\Bigl(|x_+|^2(\lambda^2+|x_+|^2){\bar x_-\dd x\wedge \dd \bar x x_-\over|x_-|^2}
+|x_-|^2(\lambda^2+|x_-|^2){\bar x_+\dd x\wedge \dd \bar x x_+\over|x_+|^2}\Bigr.    \label{k2SingularF}\\
&\qquad\Bigl.-\lambda^2(\bar x_+\dd x\wedge \dd \bar x x_-+\bar x_-\dd x\wedge \dd \bar x x_+)\Bigr)\;,\nn
\end{align}
where $x_\pm=x\pm a$. For convenience, we take $a\in\RR$ (by an $SO(4)$ rotation).
A clue about the behaviour under an inversion is obtained by looking at the prefactor, governed by the function 
\begin{align}
f_{a,\lambda}(x)=|x_+|^2|x_-|^2+\lambda^2|x_+|^2+\lambda^2|x_-|^2
\end{align}
appearing in the denominator. Under an inversion $x'=x^{-1}$, we have
\begin{align}
{1\over a^2|x|^2}f_{a,\lambda}(x)={1\over a'^2|x'|^2}f_{a',\lambda'}(x')\;,
\end{align} 
where
\begin{align}
a'&={1\over\sqrt{a^2+2\lambda^2}}\;,\nn\\
\lambda'&={\lambda\over a\sqrt{a^2+2\lambda^2}}\;.   \label{aLambdaTransf}
\end{align}
In principle, it remains to be checked that the full solution transforms like this, but it is the only possibility. 
Note that ${\lambda'\over a'}={\lambda\over a}$, so a solution with two ``well separated'' instantons remains well separated viewed from the antipode (but see below). Solutions with $a^2(a^2+2\lambda^2)=1$ are invariant under inversion.
In the geometric moduli space, it is sufficient to include sizes $0<\lambda^2\leq{1-a^4\over2a^2}$, to the left of the curve in Figure
\ref{k2Figure}.

We will now examine the behaviour of the instanton scalar $\II={1\over8\pi^2}g^{mp}g^{nq}F_{mn}{}^iF_{pq}^i$. 
Since, in the solution \eqref{k2SingularF}, $\dd x\wedge \dd \bar x$ is conjugated with different vectors in different terms, we need the more general identity
for arbitrary vectors $a,b,c,d$:
\begin{align}
\Re(\bar a e_{[a}\bar e_{b]}be^i)\Re(\bar ce_{[a}\bar e_{b]}de^i)=4\bigl[2(a\cdot c)(b\cdot d)+2(a\cdot d)(b\cdot c)-(a\cdot b)(c\cdot d)\bigr]\;,
\end{align}
where $(u\cdot v)=\Re(\bar u v)$ is the ordinary scalar product.
A partial result is
\begin{align}
&4\pi^2\sqrt g \II={8\lambda^4\over(|x_+|^2|x_-|^2+\lambda^2|x_+|^2+\lambda^2|x_-|^2)^4}\nn\\
&\quad\times\Bigl[3|x_+|^4(\lambda^2+|x_+|^2)^2+3|x_-|^4(\lambda^2+|x_-|^2)^2
	+4\lambda^2(2|x_+|^2|x_-|^2+(x_+\cdot x_-)^2)\nn\\
&\qquad+2(\lambda^2+|x_+|^2)(\lambda^2+|x_-|^2)(4(x_+\cdot x_-)^2-|x_+|^2|x_-|^2)        \label{PartialResult}\\
&\qquad-12\lambda^2|x_+|^2(\lambda^2+|x_+|^2)(x_+\cdot x_-)
              -12\lambda^2|x_-|^2(\lambda^2+|x_-|^2)(x_+\cdot x_-)
	\Bigr]\;.\nn
\end{align}
We then insert $|x_\pm|^2=|x|^2\pm2(a\cdot x)+a^2$, $(x_+\cdot x_-)=|x|^2-a^2$. Let us call the object within square brackets in eq. 
\eqref{PartialResult} $X_{a,\lambda}(x)$. Also, let $f_{a,\lambda}(x)=|x_+|^2|x_-|^2+\lambda^2|x_+|^2+\lambda^2|x_-|^2$. Then,
\begin{align}
X_{a,\lambda}(x)&=12|x|^8+16a^2|x|^6+8a^2(a^2-2\lambda^2)|x|^4\nn\\
&+16a^4(a^2+2\lambda^2)|x|^2+12a^4(a^2+2\lambda^2)^2       \label{NotSoPartialR}\\
&+\left(128|x|^4+64(6a^2+\lambda^2)|x|^2+128 a^2(a^2+2\lambda^2)\right)(a\cdot x)^2
+64(a\cdot x)^4\;, \nn
\end{align}
and
\begin{align}
f_{a,\lambda}(x)=|x|^4+2(a^2+\lambda^2)|x|^2+a^2(a^2+2\lambda^2)-4(a\cdot x)^2\;.
\label{eq:Ricci_equal_size_3}
\end{align}
It is then straightforward to verify the behaviour under inversion
\begin{align}
f_{a',\lambda'}(x^{-1})&=a^{-2}(a^2+2\lambda^2)^{-1}|x|^{-4}f_{a,\lambda}(x)\;,\nn\\
X_{a',\lambda'}(x^{-1})&=a^{-4}(a^2+2\lambda^2)^{-2}|x|^{-8}X_{a,\lambda}(x)\;.
\end{align}
Together with the second eq. in \eqref{aLambdaTransf} and $\sqrt g\mapsto |x|^8\sqrt g$, this shows that
$4\pi^2\II={8\lambda^4\over\sqrt g}{X_{a,\lambda}(x)\over f_{a,\lambda}(x)^4}$ is invariant under an inversion.
This is a good consistency check on the calculations leading to eq. \eqref{NotSoPartialR}.

We can now start to investigate the behaviour of $\II$ in (for example) the left region of Figure \ref{k2Figure}.
The values at $x=0$ and $x=\infty$ provide one interesting piece of  input:
\begin{align}
4\pi^2\II(0)&={6\lambda^4\over a^4(a^2+2\lambda^2)^2}\;,\nn\\
4\pi^2\II(\infty)&=6\lambda^4\;.
\end{align}
This gives the simple characterisation of the left half of the ``phase diagram'', Figure \ref{k2Figure}, that it consists of the solutions with
$\II(0)\geq\II(\infty)$.

If we consider $\II$ as a function of the two variables $|x|^2$ and $(a\cdot x)^2$, we note that $X_{a,\lambda}$ increases with increasing $(a\cdot x)^2$ for constant $|x|^2$, while $f_{a,\lambda}$ decreases. This implies that any local maximum must lie on the real line.

\begin{figure}
\begin{center}
\includegraphics[scale=.5]{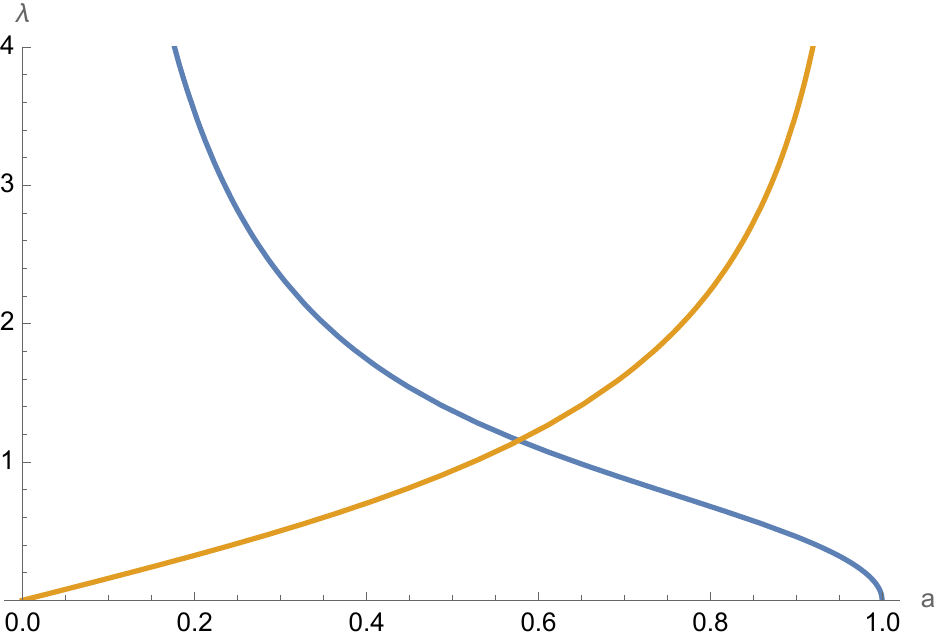}
\end{center}
\caption{The subspace of the geometric moduli space for equal size instantons, with the critical line for appearance/disappearance of twin peaks. The blue line is the same as the one in Figure \ref{k2Figure}, while the orange line divides solutions with a single peak from those that show two peaks. The intersection of the two lines occurs $a={1\over\sqrt3}$, $\lambda={2\over\sqrt3}$, which is the ``special" point of the moduli space that we will focus on.  \label{k2Figure2}}
\end{figure}

It is straightforward to see that all partial derivatives $\*_m\II$ vanish at $x=0$.
We may ask if $x=0$ is a local maximum, minimum or a saddle point. 
It turns out that a second directional derivative orthogonal to $a$ is always (in the parameter region) negative.
The second directional derivative along  $a$ may be positive or negative.
We find it to be positive for small $\lambda$ (and small enough $a$) and negative for large $\lambda$, the critical point being
$\lambda^2={a^2(a^2+5)\over2(1-a^2)}$. This divides the region of parameter space in two parts, one where the size parameter is small, so the instantons are separated, yielding two peaks, one where the size is large enough relative to the separation, so there is only a single peak. This second critical line is included in Figure \ref{k2Figure2}. 
The two curves intersect in the ``special'' point $a={1\over\sqrt3}$, $\lambda={2\over\sqrt3}$. 
For these values of the moduli, $\II$ is constant along the great circle through the origin and $a$.
There is in fact an enhancement of isometry at this point, and it can on good grounds be considered the ``center'' of the $k=2$ moduli space.

\begin{figure}
\begin{center}
\includegraphics[scale=.22]{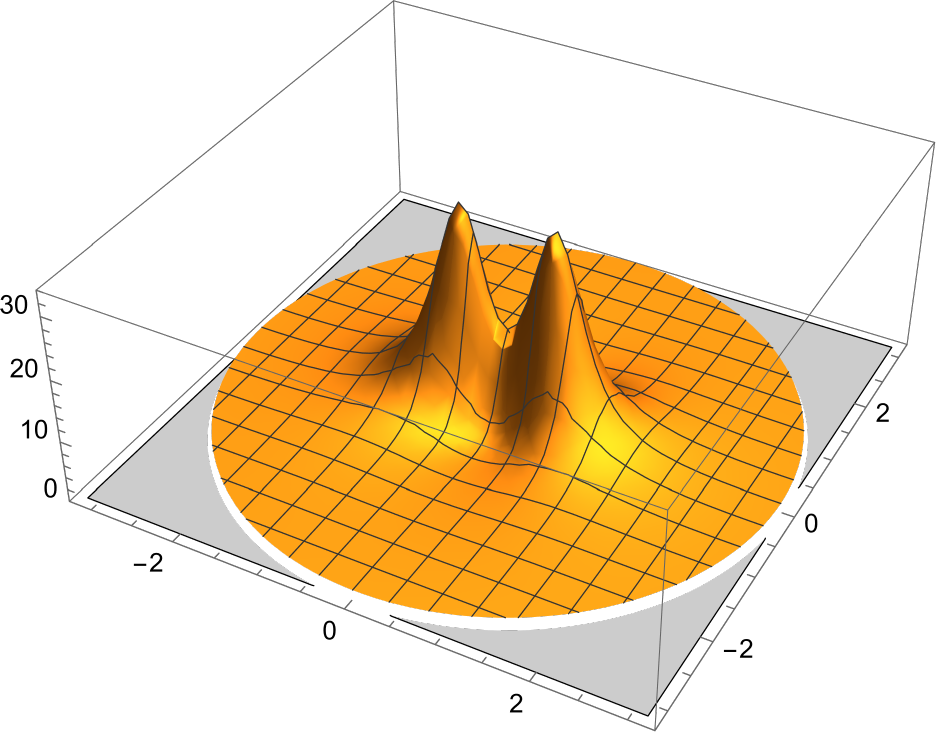}
\includegraphics[scale=.22]{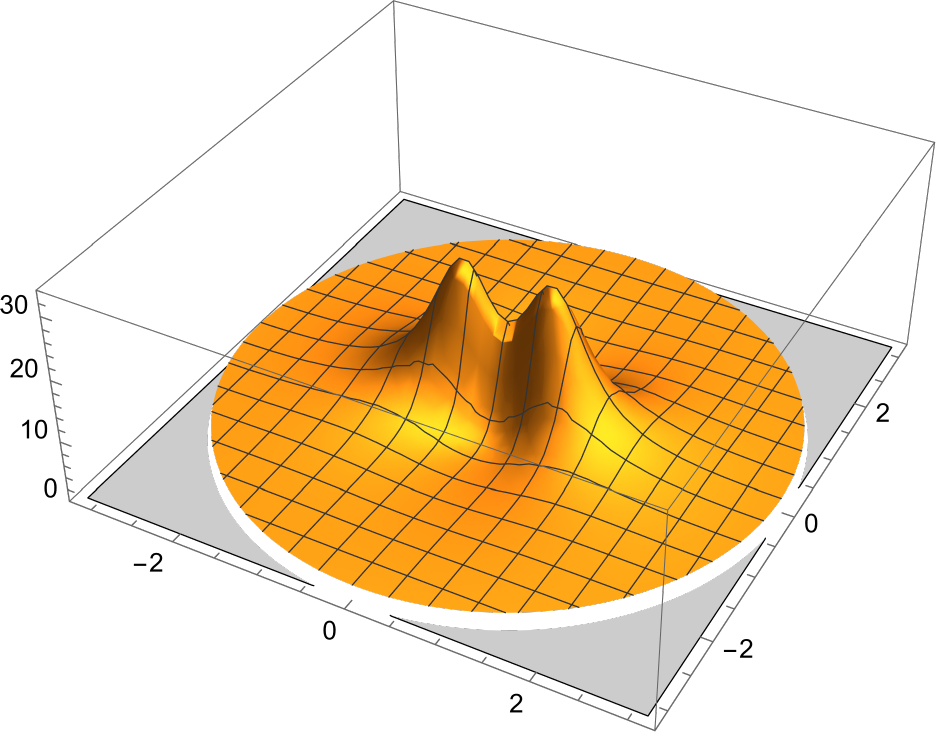}
\includegraphics[scale=.22]{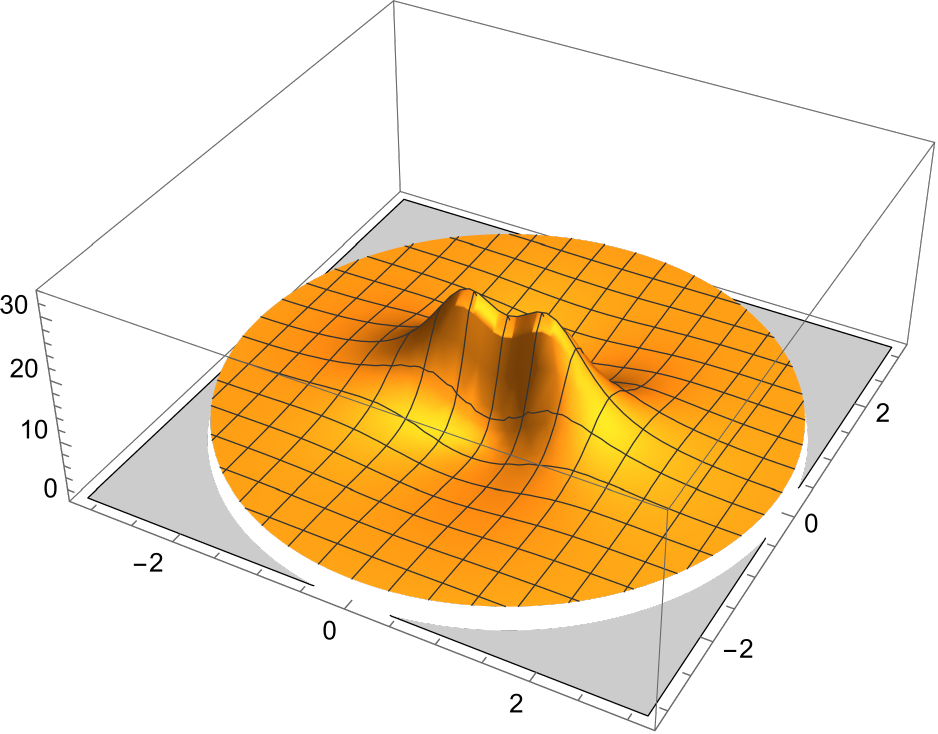}
\includegraphics[scale=.22]{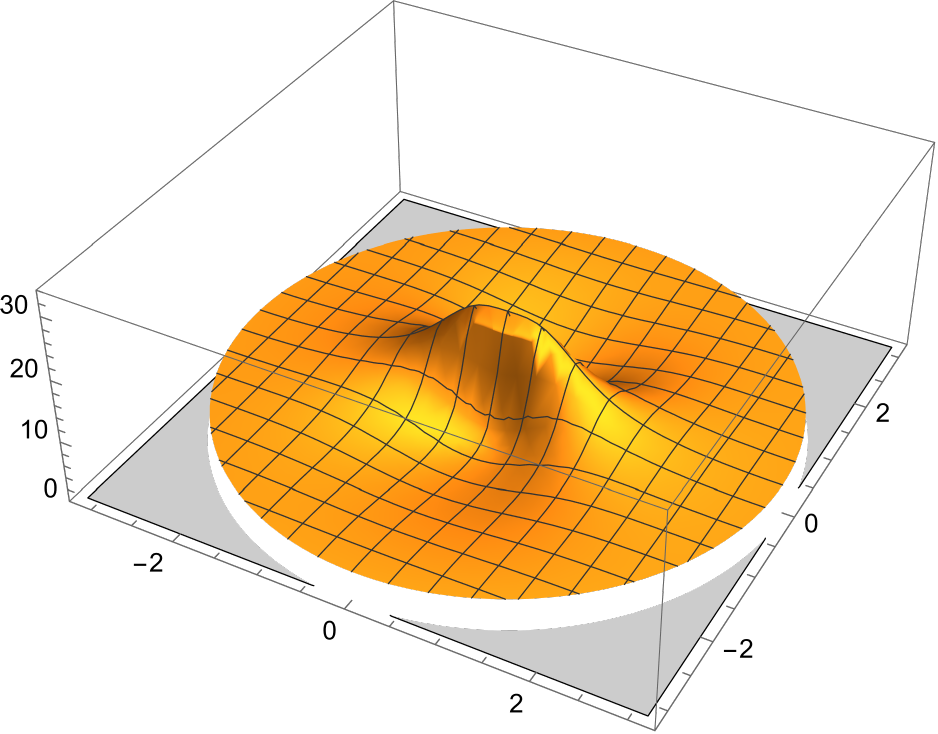}
\includegraphics[scale=.22]{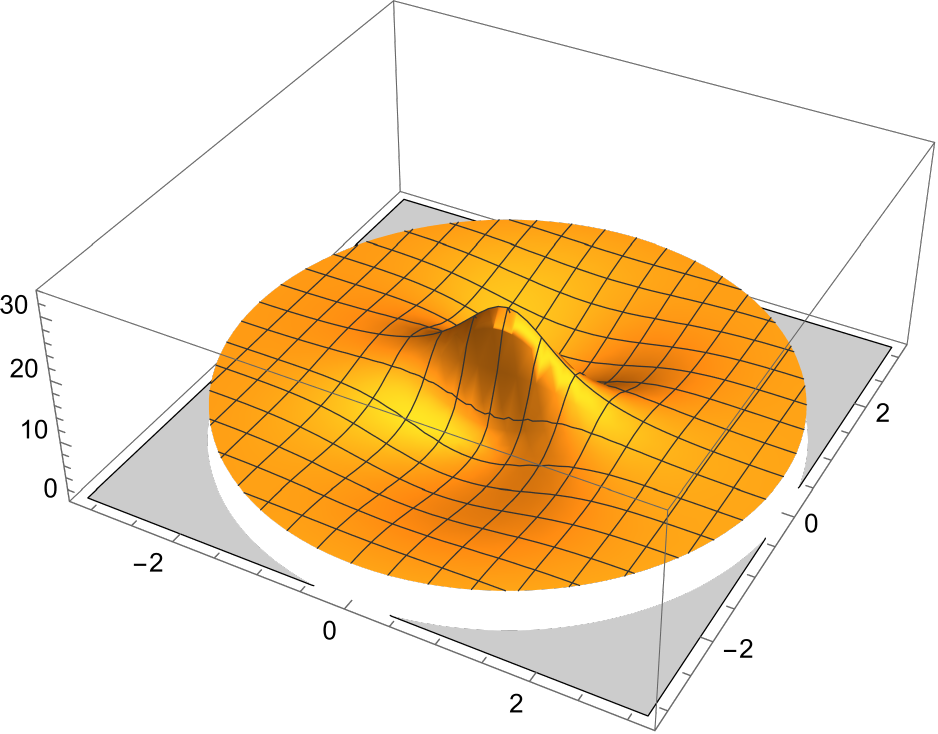}
\includegraphics[scale=.22]{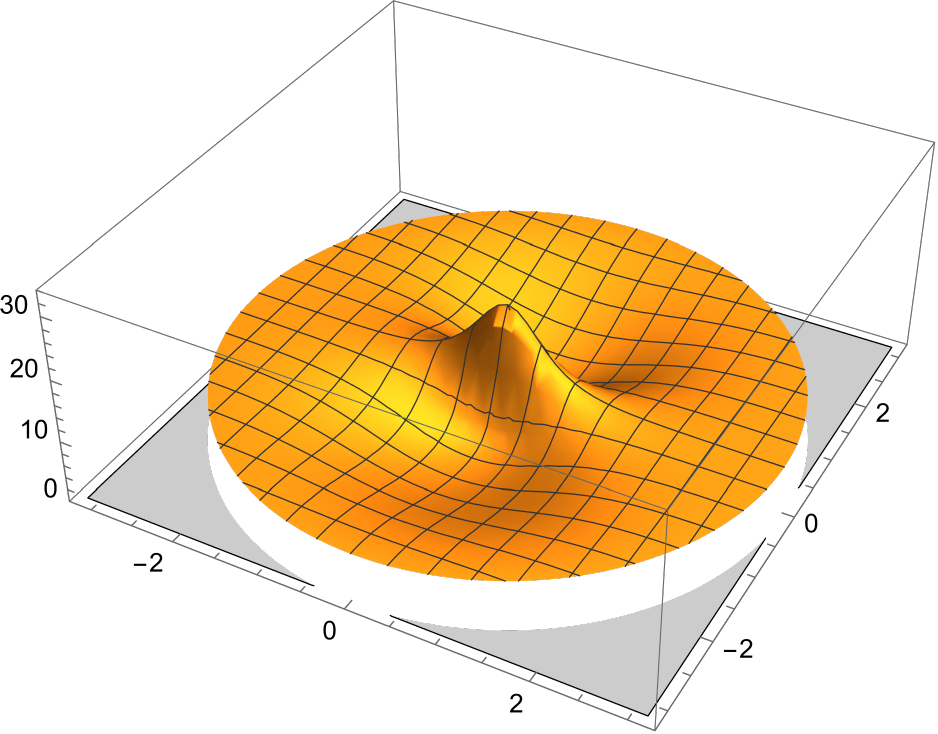}
\includegraphics[scale=.22]{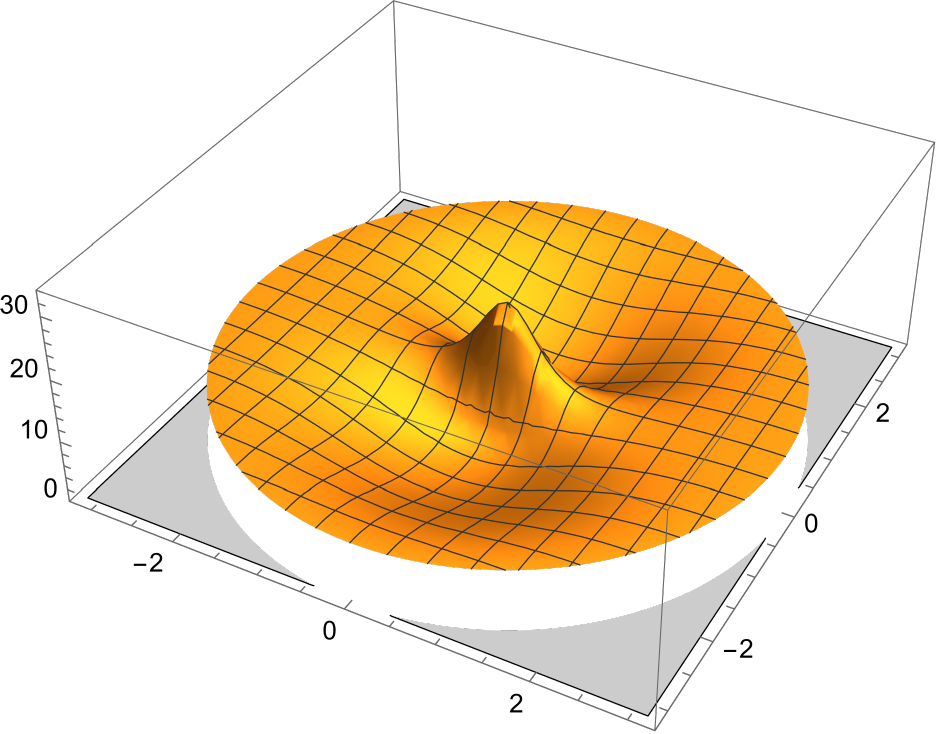}
\includegraphics[scale=.22]{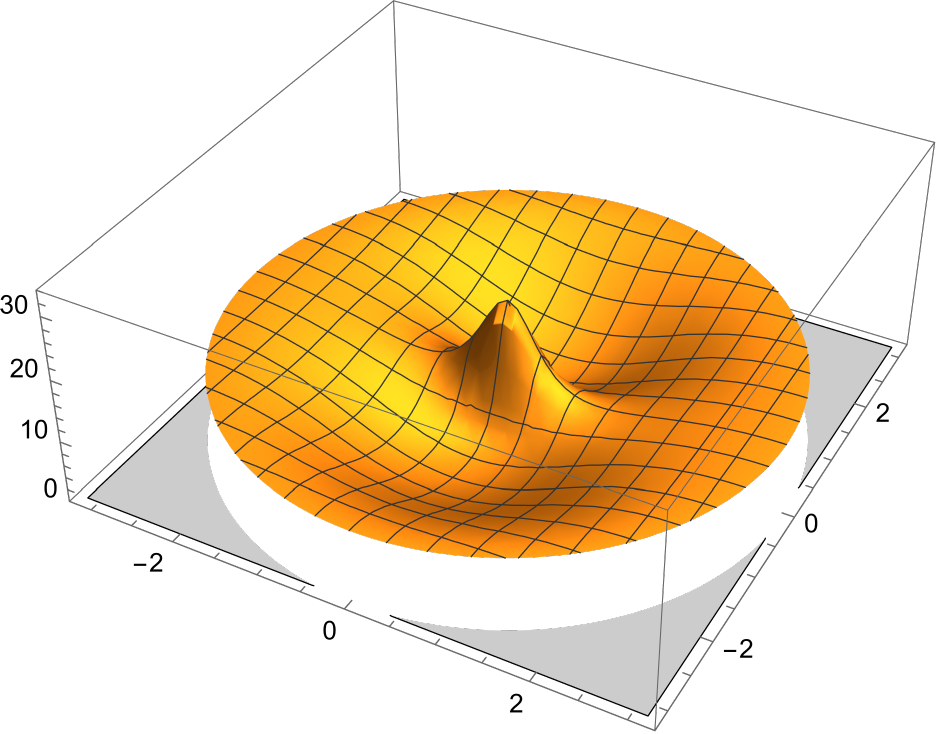}
\includegraphics[scale=.22]{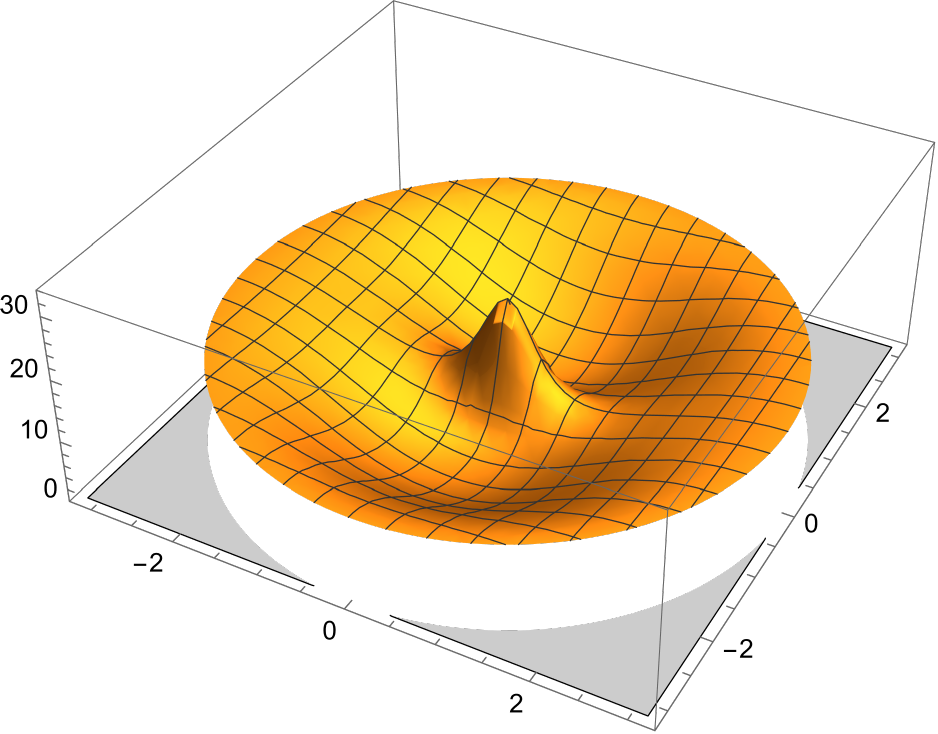}
\includegraphics[scale=.22]{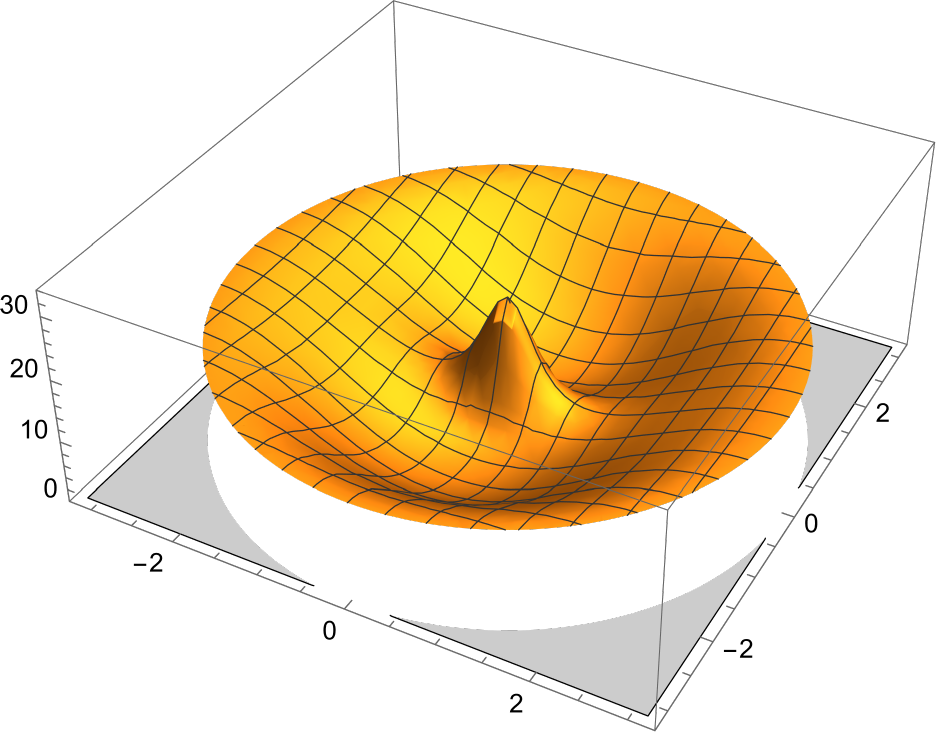}
\includegraphics[scale=.22]{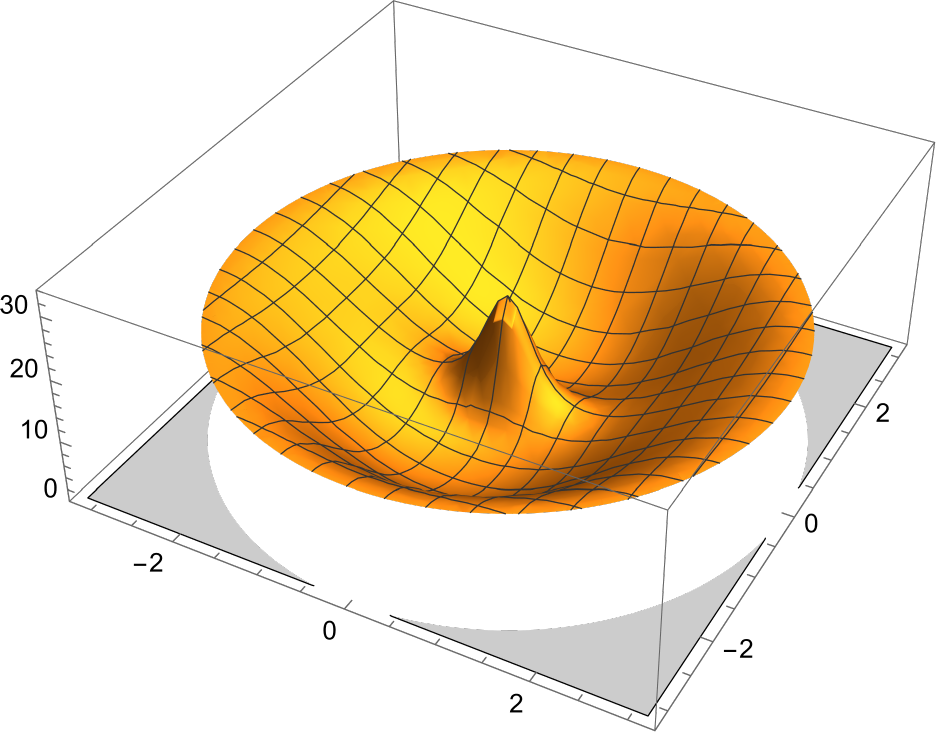}
\includegraphics[scale=.22]{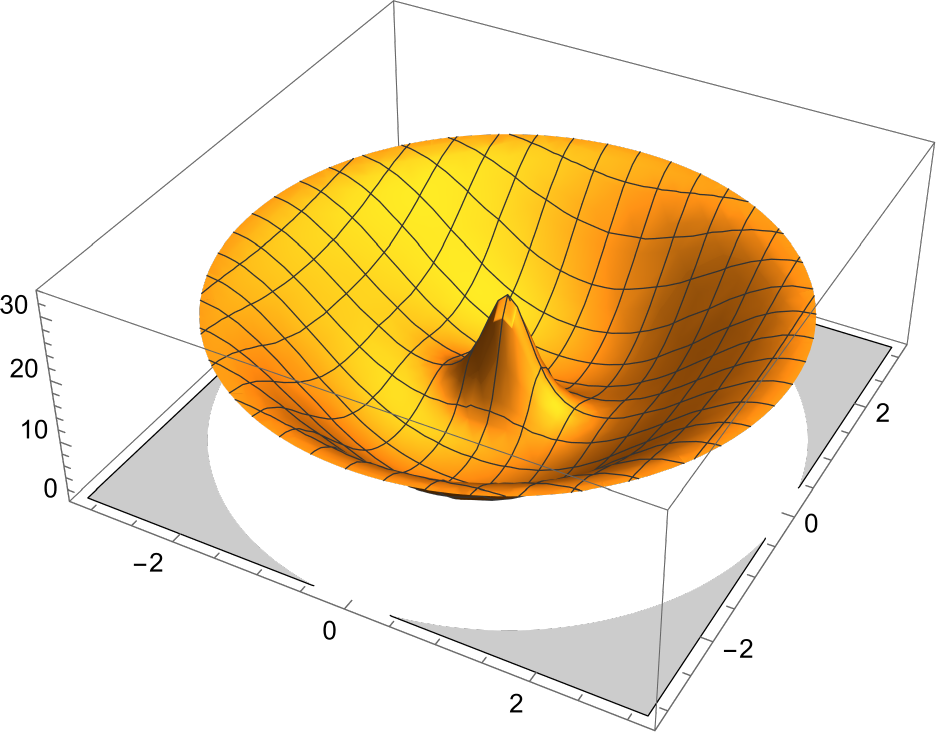}
\end{center}
\caption{Plots of $4\pi^2\II$ for $a={1\over2}$ and $\lambda^2={n\over8}$, $n=4,\dots,15$. 
The radial direction in the plot is polar angle (``$\theta$'') on the sphere.
$n=7$, the upper right plot, is on the critical line where the peaks coalesce. $n=15$, the lower right plot, is inversion-invariant.}
\label{ahalfPlots}
\end{figure}

The algebraic equation for stationary points of $\II$ at the real axis, away from $x=0$, is a cubic equation for $(\Re x)^2$. 
A careful analysis of this equation (discriminant, sum and product of roots) for all values of the parameters, gives at hand that there are no other local maxima than the ones already mentioned. The behaviour described above is illustrated in Figure \ref{ahalfPlots}, which considers multiple values of $\lambda$ for a fixed $a$, illustrating how the two peaks merge into a single one when the size becomes large enough compared to the separation, as well as the interplay between $\II(0)$ and $\II(\infty)$.

If we restrict our attention to solutions with $\II(0)\geq\II(\infty)$ (to the left of the blue curve in Figure \ref{k2Figure2}), the maximum value is $\II(0)$, as long as we are above the critical curve (orange). Below the critical curve, the maximum value is attained at the two peaks. Their precise location requires solving a cubic equation. When $\lambda\rightarrow0$, they approach $\pm a$, and their height diverges as ${3\over2}(1+a^2)^4\lambda^{-4}$.

\subsection{The special point in the $k=2$ moduli space}
\label{sec:Special_point_k=2}

\begin{figure}
\begin{center}
\includegraphics[scale=.6]{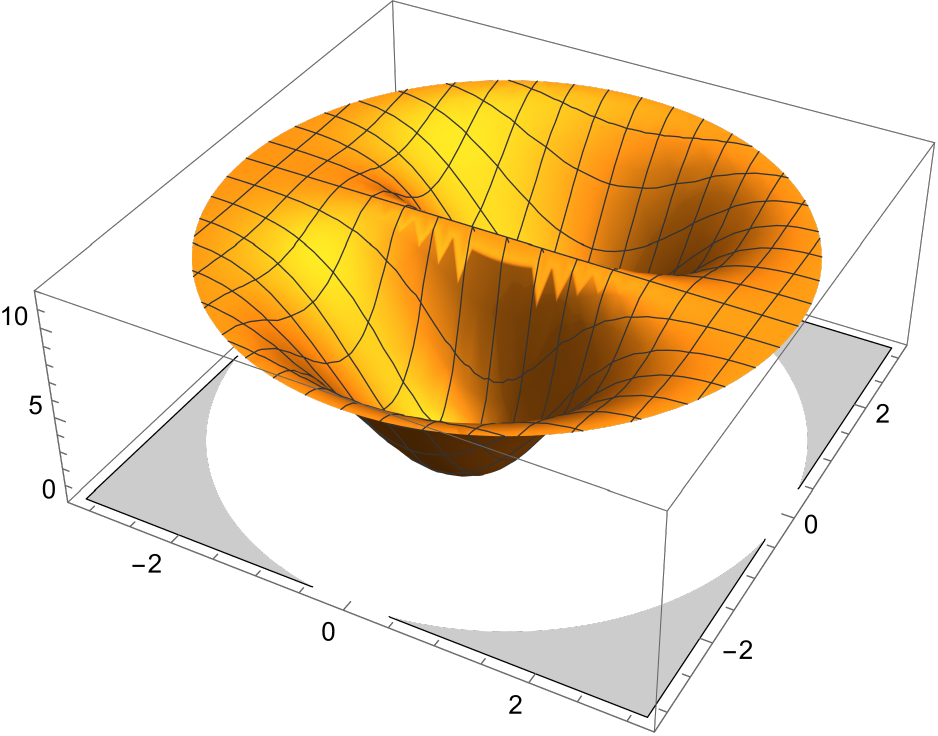}
\end{center}
\caption{Plot of $4\pi^2\II$ for $a={1\over\sqrt3}$, $\lambda={2\over\sqrt3}$. 
}
\label{specialPlot}
\end{figure}

At the special point $a={1\over\sqrt3}$, $\lambda={2\over\sqrt3}$, the  functions $f$ and $X$ can be rewritten as
\begin{align}
f_{{1\over\sqrt3},{2\over\sqrt3}}(x)&=(1+|x|^2)^2+{4\over3}|\Im x|^2\;,\\
X_{{1\over\sqrt3},{2\over\sqrt3}}(x)&=12\bigl((1+|x|^2)^4-{32\over9}(1+|x|^2)^2|\Im x|^2+{16\over27}|\Im x|^4\bigr)\;.\nn
\end{align}
The instanton scalar becomes
\begin{align}
4\pi^2\II={32\over3}{1-{32\over9}{|\Im x|^2\over (1+|x|^2)^2}+{16\over27}{|\Im x|^4\over (1+|x|^2)^4}
\over(1+{4\over3}{|\Im x|^2\over (1+|x|^2)^2})^4 } \, .
\label{SpecialI}
\end{align}
$\II$ is constant on surfaces $|\Im x|={\rho\over 2}(1+|x|^2)$. This is the stereographic image\footnote{The stereographic projection is along lines in $\RR^5=\HH\oplus\RR$ from $(0,2)$ through the point $(u,v)$: $|u|^2+(v-1)^2=1$ on a unit $S^4$ to $(2x,0)$. The factor $2$ is to obtain the standard metric $ds^2={4|\dd x|^2\over(1+|x|^2)^2}$.}
 of the space $|\Im u|=\rho$ in the unit $S^4$. The parameter $\rho$ lies in the interval $0\leq \rho\leq 1$. For $0<\rho<1$ this is 
 $S^2_\rho\times S^1_{\scaleto{\sqrt{1-\rho^2}}{8pt}}$, where the subscripts indicate radius.
 For $\rho=0$ it degenerates to $S^1$ (the compactified real line $\Im x=0$), and for $\rho=1$ to $S^2$ ($\Re x=0$, $|\Im x|=1$); this is shown in Figure \ref{specialPlot}, where only two dimension are depicted. As a consequence, the $S^2$ is represented by $S^0=\{\pm1\}$, and what looks like two minima is actually a $2$-sphere of minima.

 The maximum of $\II$ is attained at $\Im x=0$ (and $|x|=\infty$), with $4\pi^2\II={32\over3}$, and the minimum at $\Re x=0$, $|\Im x|=1$, with
 $4\pi^2\II={1\over2}$.
 
 As a check of normalisation, we can perform the integration of the instanton density for the special solution. Using the slicing in 
 $S^2_\rho\times S^1_{\scaleto{\sqrt{1-\rho^2}}{8pt}}$, we get the integration measure 
 $d^4x\sqrt g=dV_{S^4}={d\rho\over\sqrt{1-\rho^2}}dV_{S^1_{\scaleto{\sqrt{1-\rho^2}}{6pt}}}dV_{S^2_\rho}$.
 For a function which only depends on $\rho$,
 \begin{align}
 \int d^4x\sqrt g f(\rho)=8\pi^2\int_0^1 d\rho\,\rho^2f(\rho) \, ,
 \end{align}
(reproducing $\Vol(S^4)={8\pi^2\over3}$).
 Applied on $\II={8\over3\pi^2}{1-{8\over9}\rho^2+{1\over27}\rho^4\over(1+{1\over3}\rho^2)^4}$, this gives the instanton number
 \begin{align}
k= \int_{S^4}d^4x\sqrt g\II= 8\pi^2\times{8\over3\pi^2}\times{3\over32}=2\;.\label{SpecialK}
 \end{align}
 The integrals corresponding to  the three terms in the numerator each also contains a contribution to the last factor in eq. \eqref{SpecialK} which is a rational numbers times $\pi\sqrt3$. These cancel in the sum, providing a strong consistency check.

\section{Metric of maximal isometry and its Ricci tensor \label{CalcRicciSec}}
In this section, we put together all of the results accumulated so far. By focussing on the special points in the moduli space of the $k=1$ and $k=2$ instantons described in the previous section, we show that the resulting Kaluza--Klein metric has the maximal isometry, \ie, $SO(3) \times O(2)$ \cite{10.2307/1971078}, and establish bounds on the radius of the base $S^4$ to ensure positivity of the Ricci tensor.

\subsection{Special point and symmetry enhancement}

The (or, a) regular field strength $F'$ is obtained by applying a singular gauge transformation as $F'=gFg^{-1}$, where 
$g={\bar x_a^{-1}-\bar x_b^{-1}\over|\bar x_a^{-1}-\bar x_b^{-1}|}$. Focussing on the special values of the moduli discussed in Section \ref{sec:Special_point_k=2}, a calculation yields:
\begin{align}
F'&={4/3\over((1+|x|^2)^2+{4\over3}|\Im x|^2)^2}\label{regk2sol}\\
&\times\left(Q_0 \dd x\wedge \dd \bar x
+Q_1 ({\Im x\over|\Im x|} \dd x\wedge \dd \bar x-\dd x\wedge \dd \bar x{\Im x\over|\Im x|})
+Q_2{\Im x\over|\Im x|}\dd x\wedge \dd \bar x {\Im x\over|\Im x|}\right)\;,\nn
\end{align}
where
\begin{align}
Q_0&=2(1+|x|^2)^2-{2\over 3}(5+3|x|^2)|\Im x|^2\;,\nn\\
Q_1&=2(1+|x|^2)\Re x|\Im x|\;,\\
Q_2&=-{2\over3}(1+3|x|^2)|\Im x|^2\;.\nn
\end{align}
It is however more convenient to use an ``orthogonal'' set $\{\omega_I\}_{I=1}^3$ for the selfdual $\su(2)$-valued 2-forms:
\begin{align}
\omega_1&={1\over8}\Bigl({\Im x\over|\Im x|} \dd x\wedge \dd \bar x-\dd x\wedge \dd \bar x{\Im x\over|\Im x|}\Bigr)\;,\nn\\
\omega_2&={1\over8}\Bigl(\dd x\wedge \dd \bar x+{\Im x\over|\Im x|}\dd x\wedge \dd \bar x {\Im x\over|\Im x|}\Bigr)\;,\\
\omega_3&={1\over8}\Bigl(\dd x\wedge \dd \bar x-{\Im x\over|\Im x|}\dd x\wedge \dd \bar x {\Im x\over|\Im x|}\Bigr)\;.\nn
\end{align}
Then, defining $M_{IJ}{}^{ij}=\sqrt gg^{mp}g^{nq}\omega_{Imn}{}^i\omega_{Jpq}{}^j$, $M_{IJ}^{ij}=0$ for $I\neq J$, and
$M_{11}{}^{ij}=M_{22}{}^{ij}=\delta^{ij}-{x^ix^j\over|\Im x|^2}\equiv \Pperp{ij}$, $M_{33}{}^{ij}={x^ix^j\over|\Im x|^2}\equiv \Ppar^{ij}$. These are the projection operators on imaginary quaternions ortogonal and parallel to $\Im x$, respectively.
A M\"obius transformation $x\mapsto(x+\beta)(1-\beta x)^{-1}$, $\beta\in\RR$, preserves ${\Im x\over1+|x|^2}$
(and thus ${\Im x\over|\Im x|}$). 
This means that this $SO(2)$ rotation 
leaves these projection operators invariant. 
It acts on $\dd x\wedge \dd \bar x$ as
\begin{align}
\dd x\wedge \dd \bar x&\mapsto (1+\beta^2)|1-\beta x|^{-4}{1-\beta x\over|1-\beta x|}\dd x\wedge \dd \bar x{1-\beta\bar x\over|1-\beta x|}\;,\nn\\
{\dd x\wedge \dd \bar x\over(1+|x|^2)^2}&\mapsto {1-\beta x\over|1-\beta x|}{\dd x\wedge \dd \bar x\over(1+|x|^2)^2}{1-\beta\bar x\over|1-\beta x|}\;.
\end{align}
The conjugation with ${1-\beta x\over|1-\beta x|}$ commutes with $\Im x$.
Thus, an $SO(2)$ rotation induces an $SU(2)$ gauge transformation. An element $h_\varphi=\cos\varphi+{\Im x\over|\Im x|}\sin\varphi$ transforms $\omega_I$ as 
$\omega_I\mapsto h_\varphi\omega_I\bar h_\varphi=(R_\varphi)_I{}^J\omega_J$, with
\begin{align}
R_\varphi=\left(\begin{matrix}\cos2\varphi&\sin2\varphi&0\\-\sin2\varphi&\cos2\varphi&0\\0&0&1\end{matrix}\right)\;.
\end{align}
This implies that $F$ is invariant modulo a gauge transformation, and $SO(2)$ is an iso\-metry. Together with the $\ZZ_2$ transformation $\Re x\mapsto -\Re x$, we obtain an $O(2)$.

Expressing $F'$ in the new basis, 
\begin{align}
F¨'={4/3\over((1+|x|^2)^2+{4\over3}|\Im x|^2)^2}q^I\omega_I\;,
\end{align}
where
\begin{align}
(q^1)^2+(q^2)^2&=64(1+|x|^2)^4\Bigl(1-4{|\Im x|^2\over(1+|x|^2)^2}\Bigr)\;,\nn\\
q^3&=8(1+|x|^2)^2\Bigl(1-{4\over3}{|\Im x|^2\over(1+|x|^2)^2}\Bigr)\;.
\end{align}
This shows that $Y^{ij}={1\over2}g^{mp}g^{nq}F'_{mn}{}^iF'_{pq}{}^j$ is invariant under $SO(2)$ and takes the form
\begin{align}
Y^{ij}&={32\over9}\Bigl(1+{4\over3}{|\Im x|^2\over (1+|x|^2)^2}\Bigr)^{-4}\nn\\
&\times\left[\Bigl(1-4{|\Im x|^2\over (1+|x|^2)^2}\Bigr)\delta^{ij}
+{4\over3}{|\Im x|^2\over (1+|x|^2)^2}\Bigl(1+{4\over 3}{|\Im x|^2\over(1+|x|^2)^2}\Bigr){x^ix^j\over|\Im x|^2}
\right]\\
&={32\over9}(1+{\rho^2\over3})^{-4}\left[(1-\rho^2)\delta^{ij}+{\rho^2\over3}(1+{\rho^2\over3}){x^ix^j\over|\Im x|^2}
\right]\;.\nn
\end{align}
Taking the trace gives back eq. \eqref{SpecialI}.
The factor $1-4{|\Im x|^2\over (1+|x|^2)^2}=1-\rho^2$ is positive semidefinite, it has its global minimum $0$ at $\Re x=0$, $|\Im x|=1$. 
$Y^{ij}$ is a positive definite combination of the projection operators $P^{ij}_{/\!/}={x^ix^j\over|\Im x|^2}$ and $P^{ij}_\perp=\delta^{ij}-{x^ix^j\over|\Im x|^2}$, with coefficients that are invariant under the ``extra'' $SO(2)$.

\subsection{The exotic sphere with maximal isometry}

The bundle vielbein is given as in Section \ref{BundleGeometry}, with 
$\sF=F-yG\bar y$, where $G$ is the regular $SO(5)$-symmetric $k=1$ solution \eqref{k1solution} with $\lambda=1$ and $a=0$ (we drop the primes), and $F$ is the regular $k=2$ solution at the special point, eq. \eqref{regk2sol}.
Notice that there is no freedom in relative positioning on $S^4$ of the $k=1$ and $k=2$ solutions, when the $k=1$ instanton is at the symmetric values of the moduli. 

The product space $S^4\times S^3$ of course has isometry $SO(5)\times SO(4)$.
Any instanton solution for the left/right $SU(2)$ of $S^3$ links the left/right isometry of $S^3$ to one of the $SU(2)$'s in 
$(SU(2)\times SU(2))/\ZZ_2\simeq SO(4)\subset SO(5)$ of $S^4$ through the 't Hooft symbols. 
The maximally symmetric $k=1$ solution has isometry $SO(5)\times SO(3)$, which at the particular value of the relative radius giving the round $S^7$ gets enhanced to $SO(8)$. The maximally symmetric $k=2$ solution has isometry $SO(3)\times SO(3)\times O(2)$.
The bundle at hand, with both left and right instantons, thus breaks all $S^3$ isometry. The remaining isometry is
$SO(3)\times O(2)$.

\subsection{Ricci tensor and its bounds}

In order to calculate the (field strength)${}^2$ contributions to the components of the Ricci tensor, let components of a field strength be real proportional to
\begin{align}
f(\alpha,\beta)={1\over2}(\bar \alpha\dd x\wedge \dd \bar x\beta+\bar \beta\dd x\wedge\dd \bar x\alpha)\;,
\end{align}
$\alpha,\beta\in\HH$,
and define the map from the tensor product of $\HH'$-valued selfdual 2-forms to $\vee^2\HH'$: $\varrho^{ij}(f,g)=\sqrt gg^{mp}g^{nq}f_{mn}{}^{(i}g_{pq}{}^{j)}$. Then,
\begin{align}
\varrho^{ij}(f(\alpha,\beta),f(\gamma,\delta))=-8\Re(e^{(i}\bar \alpha\gamma e^{j)}\bar \delta\beta)-8\Re(e^{(i}\bar \alpha\delta e^{j)}\bar \gamma\beta)\;.
\end{align}
If the field strength is conjugated by $u$, $|u|=1$, $F\mapsto uF\bar u$, we instead need to calculate
$\varrho^{ij}(f(\alpha\bar u,\beta\bar u),f(\gamma\bar u,\delta \bar u))$. This is equivalent to conjugating the $e^i$'s by $\bar u$, and is in general different from $\varrho^{ij}(f(\alpha,\beta),f(\gamma,\delta))$, unless at least 3 of the quaternions $\alpha,\beta,\gamma,\delta$ are real proportional to each other, 
making $\varrho^{ij}(f(\alpha,\beta),f(\gamma,\delta))$ proportional to $\delta^{ij}$.

When constructing the contribution from $\sF^2$ to the components $R_{ij}$ of the Ricci tensor, they will all be of the above form. Contraction with $\delta^{ij}$ gives the terms in the contribution to $R_{ii}$, but also to $R_{aa}$. We can then observe, than since the field strength is selfdual, the $\sF^2$ contribution to $R_{ab}$ is automatically proportional to $\delta_{ab}$. So, calculating 
$\varrho^{ij}(f(\alpha,\beta),f(\gamma,\delta))$ for the various terms in $\sF$ gives all information needed for the whole Ricci tensor.

We choose, in $\sF=F-yG\bar y$, to let $F$ be the 2-instanton solution. When the centra $a,b\in\RR$, the only (non-real) quaternion appearing multiplying $\dd x\wedge\dd \bar x$ from the left or right is ${\Im x\over|\Im x|}$, which is abbreviated as $I$ below. $G$ is the taken to be the 1-instanton solution, which is conjugated by $y$. The relevant $\varrho^{ij}$'s can be calculated as:
\begin{align}
\varrho^{ij}(f(1,1),f(1,1))&=16\delta^{ij}\;,\nn\\
\varrho^{ij}(f(1,1),f(1,I))&=0\;,\nn\\
\varrho^{ij}(f(1,1),f(I,I))&=-16\delta^{ij}+32I^iI^j\;,\nn\\
\varrho^{ij}(f(1,I),f(1,I))&=16(\delta^{ij}-I^iI^j)\;,\nn\\
\varrho^{ij}(f(1,I),f(I,I))&=0\;,\\
\varrho^{ij}(f(I,I),f(I,I))&=16\delta^{ij}\;,\nn\\
\varrho^{ij}(f(1,1),f(y,y))&=16\delta^{ij}(1-2|\Im y|^2)+32y^iy^j\;,\nn\\
\varrho^{ij}(f(1,I),f(y,y))&=32y_0(\delta^{ij}(y\cdot I)-y^{(i}I^{j)})+32\epsilon^{(i}{}_{kl}y^{j)}y^kI^l\;,\nn\\
\varrho^{ij}(f(I,I),f(y,y))&=-16\delta^{ij}(1-2|\Im y|^2)+32I^iI^j(1-2|\Im y|^2)\nn\\
&\qquad -32y^iy^j+64(y\cdot I)y^{(i}I^{j)}
+64y_0\epsilon^{(i}{}_{kl}I^{j)}I^ky^l\;,\nn\\
\varrho^{ij}(f(y,y),f(y,y))&=16\delta^{ij}\;.\nn
\end{align}
The first six can be obtained from the following three by letting $y=1$ or $y=I$.
It is convenient to use the linear combinations
\begin{align}
\omega_1&=-{1\over4}f(1,I)={1\over8}\Bigl(I \dd x\wedge \dd \bar x-\dd x\wedge \dd \bar xI\Bigr)\;,\nn\\
\omega_2&={1\over8}(f(1,1)-f(I,I))={1\over8}\Bigl(\dd x\wedge \dd \bar x+I\dd x\wedge \dd \bar x I\Bigr)\;,\\
\omega_3&={1\over8}(f(1,1)+f(I,I))={1\over8}\Bigl(\dd x\wedge \dd \bar x-I\dd x\wedge \dd \bar x I\Bigr)\;.\nn
\end{align}
in the expansion of the 2-instanton field strength $F$. They fulfil
$\varrho^{ij}(\omega_I,\omega_J)=0$ for $I\neq J$ and
$\varrho^{ij}(\omega_1,\omega_1)=\varrho^{ij}(\omega_2,\omega_2)=\delta^{ij}-I^iI^j=\Pperp^{ij}$,
$\varrho^{ij}(\omega_3,\omega_3)=I^iI^j=\Ppar^{ij}$, which are projection matrices on the imaginary quaternions orthogonal and parallel to $I$, respectively. 

We now let $\sF=F-yG\bar y$, with 
$F=(1+[x|^2)^{-2}q^I\omega_I$ and $G={1\over4}(1+|x]^2)^{-2}\gamma f(1,1)$.	

Then,
\begin{align}
(1+|x|^2)^4\varrho^{ij}(\sF,\sF)
&=\gamma^2\delta^{ij}-{1\over2}\gamma q^I\varrho(\omega_I,f(y,y))+((q^1)^2+(q^2)^2)\Pperp^{ij}+(q^3)^2\Ppar^{ij}\;,
\end{align}
so, with $S^4$ having unit radius,
\begin{align}
g^{mp}g^{nq}\sF_{mn}{}^i\sF_{pq}{}^j
&={1\over16}\bigl[\gamma^2\delta^{ij}-{1\over2}\gamma q^I\varrho^{ij}(\omega_I,f(y,y))\nn\\
&\quad+((q^1)^2+(q^2)^2)\Pperp^{ij}+(q^3)^2\Ppar^{ij}\bigr]\;.\label{rhoFF}
\end{align}
The mixed term is somewhat complicated. The three symmetric $(3\times3)$-matrices $\varrho(\omega_I,f(y,y))$ have entries that are functions on $S^3$. 
Let $y=\xi+\eta I+\zeta J$ in a local quaternionic basis\footnote{Notice that this basis is local both on $S^4$ ($\Im x$ defines the $I$ direction) and on $S^3$ ($\Im y$ then defines the $IJ$ plane), and in general not used for anything but local algebraic considerations.} $(1,I,J,K)$, where $J={y-(y\cdot I)I\over|y-(y\cdot I)I|}$ and $K=IJ$ (the basis degenerates if $\Im y$ is parallel to $I$, but that case is easy to treat). 
The coefficients obey $\xi^2+\eta^2+\zeta^2=1$.
Then,
\begin{align}
{1\over4}f(y,y)&=-2\xi\eta\omega_1+(\xi^2-\eta^2)\omega_2+(\xi^2+\eta^2)\omega_3\nn\\
&+{1\over2}\xi\zeta f(1,J)+{1\over2} \eta\zeta f(I,J)+{1\over4}\zeta^2f(J,J)\;.
\end{align}
We can now calculate the three matrices $M_I^{ij}={1\over4}\varrho(\omega_I,f(y,y))$ occurring in the mixed term. In the $IJK$ basis they are
\begin{align}
M_1&=\left(\begin{matrix}\qquad0\qquad&\xi\zeta&\eta\zeta\\\xi\zeta&\quad-2\xi\eta\quad&\zeta^2
			\\\eta\zeta&\zeta^2&\quad-2\xi\eta\quad\end{matrix}\right)\;,\nn\\
M_2&=\left(\begin{matrix}\qquad0\qquad&-\eta\zeta&-\xi\zeta\\-\eta\zeta&\xi^2-\eta^2+\zeta^2&\qquad0\qquad\\
			-\xi\zeta&\qquad0\qquad&\quad \xi^2-\eta^2-\zeta^2\quad\end{matrix}\right)\;,\\
M_3&=\left(\begin{matrix}\quad \xi^2+\eta^2-\zeta^2&-\eta\zeta&\xi\zeta\\
			-\eta\zeta&\qquad0\qquad&\qquad0\qquad\\\xi\zeta&0&0\end{matrix}\right)\;.\nn
\end{align}
It turns out that all eigenvalues of all three matrices lie in the interval $[-1,1]$ everywhere on $S^3$.
Inserting in eq. \eqref{rhoFF},
\begin{align}
Y^{ij}&\equiv{1\over2}g^{mp}g^{nq}\sF_{mn}{}^i\sF_{pq}{}^j\nn\\
&={1\over32}\bigl(\gamma^2\delta^{ij}-2\gamma q^I M_I^{ij}
+((q^1)^2+(q^2)^2)\Pperp^{ij}+(q^3)^2\Ppar^{ij}\bigr)\;.
\label{FFfunc}
\end{align}
All the $y$-dependence is in the matrices $M_I$. 

We are interested in finding bounds of the eigenvalues of this matrix.
In principle, this can be done by solving the cubic equations for the eigenvalues and study their dependence on $y$ and on the components $q^I$ (which depend on $x$). In practise, we only want to solve those cubic equations that reduce to quadratic ones.
We also want to use some properties of the solution at the special point in the $k=2$ moduli space.

We saw that we could freely rotate between $\omega_1$ and $\omega_2$. We use that freedom to set $q^1=0$ (this is a gauge choice; the contribution to $R_{ij}$ changes under gauge transformations).
It turns out that it is practical to use the basis $M_\pm=M_3\pm M_2$ (which means going back to $f(1,1)$ and $f(I,I)$).
Eq. \eqref{FFfunc} then becomes
\begin{align}
Y^{ij}={1\over32}\bigl(\gamma^2\delta^{ij}-2\gamma q^+ M_+^{ij}-2\gamma q^- M_-^{ij}
+(q^++q^-)^2\Ppar+(q^+-q^-)^2\Pperp   \bigr)\;.
\label{FFfunc1}
\end{align}
The eigenvalue structure of $M_\pm$ is quite simple.
The eigenvalues of $M_+$ are $\{-1+2\xi^2,-1+2\xi^2,1\}$ with eigenvectors
$\{ (0,0,1),(\zeta,\eta,0),(-\eta,\zeta,0) \}$
The eigenvalues of $M_-$ are $\{-1+2\eta^2,-1+2\eta^2,1\}$ with eigenvectors
$\{ (0,1,0),(-\zeta,0,\xi),(\xi,0,\zeta) \}$.
In order to give a lower bound on eq. \eqref{FFfunc1}, we want to complete the square to absorb negative terms. This only needs to be done for positive eigenvalues of $M_\pm$. Three regions of $S^3$ need to be considered:
\begin{itemize}
\item $\xi^2\leq{1\over2}$, $\eta^2\leq{1\over2}$. Both $M_+$ and $M_-$ have $1$ as the only positive eigenvalue.
\item $\xi^2> {1\over2}$. All eigenvalues of $M_+$ are positive, $1$ is the only positive eigenvalue of $M_-$.
\item $\eta^2>{1\over2}$. All eigenvalues of $M_-$ are positive, $1$ is the only positive eigenvalue of $M_+$.
\end{itemize}

Denote the projections on the eigenvalue $1$ subspaces of $M_\pm$ as $\Pi_\pm$.
In the first region, we write
\begin{align}
32Y&=(\gamma-q^+\Pi_+-q^-\Pi_-)^2-2\gamma q^+(M_+-\Pi_+)-2\gamma q^-(M_--\Pi_-)\nn\\
&\quad-(q^+\Pi_++q^-\Pi_-)^2
+(q^++q^-)^2\Ppar+(q^+-q^-)^2\Pperp\;.
\end{align}
All terms on the first line are non-negative, as are the terms $(q^+)^2(1-\Pi_+)+(q^-)^2(1-\Pi_-)$, so we have
\begin{align}
32Y\geq q^+q^-(-2+4\Ppar-\Pi_+\Pi_--\Pi_-\Pi_+)\label{qpqmlimit}
\end{align}
(inequality between matrices meaning contracted with any vector as $v^\transpose Yv$).
Solving for the eigenvalues of this matrix involves a ``hard''  cubic equation. 
Instead we discard the positive term with $\Ppar$. The maximal eigenvalue of $\Pi_+\Pi_-+\Pi_-\Pi_+$ takes the maximum value $2$ in the region 
(when $\zeta=0$, \ie, at $(\xi,\eta,\zeta)=({1\over\sqrt2},{1\over\sqrt2},0)$), so $32Y\geq-4q^+q^-$.
For the special solution, as a function of $\rho={2|\Im x|\over1+|x|^2}$, $q^+q^-$ takes its maximal value $4$ at $\rho=1$, so the result from the first region is
\begin{align}
Y\geq-{1\over2}\;.\label{Ylimit}
\end{align}
The same procedure in the other two regions yields the same limit, as does completing the square with the whole matrix
$q^+M_++q^-M_-$, disregarding eigenvalue signs.

If we apply this to the Ricci tensor on the exotic $S^7$, and let $S^4$ have radius $r$,
\begin{align}
R_{ij}=2\delta_{ij}+{1\over2r^4}Y_{ij}\;.
\end{align} 
Inserting the limit \eqref{Ylimit} shows that $R_{ij}$ is positive definite when $r^4>{1\over8}$ ($r\gtrapprox0.5946$).

In order to get the contribution to $R_{ab}$, we use 
$\sF_a{}^{ci}\sF_{bc}^i={1\over4}\delta_{ab}\sF^{cdi}\sF_{cd}{}^i={1\over2}\delta_{ab}Y^{ii}$ (the product of two selfdual 2-forms does not contain a traceless symmetric tensor). We immediately get
\begin{align}
32Y^{ii}=3\gamma^2-2\gamma q^+(4\xi^2-1)-2\gamma q^-(4\eta^2-1)
+(q^++q^-)^2+2(q^+-q^-)^2\;.
\end{align}
The $\gamma^2$ term from $G^2$ is proportional to the 1-instanton density and the $q^2$ terms from $F^2$ to the 2-instanton density. The mixed $\gamma q$ terms from $FG$ have average\footnote{On a unit $S^n$, the average value of the square of a coordinate in the embedding $\RR^{n+1}$ is ${1\over n+1}$.} 0 over $S^3$, so
${1\over8\pi^2}\int_{S^4}d^4x\sqrt gg^{mp}g^{nq}\sF_{mn}{}^i\sF_{pq}{}^i=3$.
For the special solution, the maximal value is attained at $\Im x=0$ and $\xi=\eta=0$ (\ie, $\Re y=0$, $\Re(\bar xy)=0$), and is
$Y^{ii}_{\hbox{\tiny max}}={3\over2}+{8\over3}+{32\over3}={89\over6}$, the three terms representing $G^2$, $FG$ and $F^2$, respectively.
For $S^4$ of radius $r$, 
\begin{align}
R_{ab}=\delta_{ab}\bigl({3\over r^2}-{1\over4r^4}Y^{ii}\bigr)\geq\delta_{ab}\bigl({3\over r^2}-{89\over24r^4}\bigr)\;.
\label{eq:Bound_R_ab}
\end{align}
$R_{ab}$ is positive definite when $r^2>{89\over72}$ ($r\gtrapprox1.112$). This limit is stronger than the one obtained from positivity of $R_{ij}$.

\section{Energy conditions}
\label{sec:Energy_conditions}

In this section, we finally come to applying our results in a physical setting, to determine whether a static exotic sphere solution defines a physically acceptable space-time. To do so, we consider some of the energy conditions. Assuming a mostly plus signature, so that $v^A v_A < 0$ defines a \textit{time-like} vector $v^A$, the most famous ones read (\cite{Curiel:2014zba,Martin-Moruno:2017exc}):
\begin{itemize}
    \item Weak Energy Condition (WEC): $G_{AB} v^A v^B \geq 0 $ for $v^A$ time-like.
     \item Strong Energy Condition (SEC): $(T_{AB} - \frac{1}{2} T g_{AB}) v^A v^B \geq 0 \iff R_{AB} v^A v^B \geq 0$ for $v^A$ time-like.
     \item Null Energy Condition (NEC): $T_{AB} k^A k^B \geq 0 \iff G_{AB} k^A k^B \geq 0$ for $k^A$ null.
     \item Dominant Energy Condition (DEC): $ G_{AB} v^A v^B \geq 0$ for $v^A$ time-like and $-G^A{}_B v^B$ is causal.
\end{itemize}
In the above, $T_{AB}$ is the stress-energy tensor, which equals (modulo Einstein gravitational constant) the Einstein tensor $G_{AB}$. The SEC and WEC both imply the NEC, while there in general is no implication between them. 

For a static space-time without any warp factor, a sufficient condition to satisfy the SEC is to have a spatial manifold with non-negative Ricci curvature (see \cite{Econditions}, for instance). This, however, is automatically true provided that the bound after \eqref{eq:Bound_R_ab} is met. 
For a space-time with scalar curvature $R\geq0$, $G_{AB}v^Av^B=R_{AB}v^Av^B-{1\over2}Rv^2\geq R_{AB}v^Av^B$ if $v^2\leq0$, so SEC implies WEC.

The dominant energy condition states that, for any future-pointing vector $v$ with $v^2\leq0$, the vector $w^A=-G^A{}_Bv^B$ also satisfies the same condition. For a static space-time with non-negative Ricci tensor this is also automatically satisfied. 
Namely, the vector $w$ becomes $w^0={1\over2}Rv^0$, $w^m={1\over2}Rv^m-R^m{}_nv^n$, and
\begin{align}
w^2={1\over4}R^2v^2-(RR_{mn}-R_{mp}R^p{}_n)v^mv^n\;.
\end{align}
The matrix $RR_{mn}-R_{mp}R^p{}_n$ is easily seen to have non-negative eigenvalues if $R_{mn}$ has non-negative eigenvalues, so
$w^2\leq{1\over4}R^2v^2\leq0$. Hence, we see that, provided $r \geq \frac{89}{72}$, all four physical energy conditions are met: weak, strong, null and dominant.

\section{Conclusions}
\label{sec:Conclusions}

In this paper, we have focused on metrics of the Kaluza--Klein type defined on the Gromoll--Meyer sphere, which is one of the exotic $7$-spheres. The ingredients of this construction are: round metrics on $S^4$ (the base space) and $S^3$ (the fibre), and $k=1,2$ $SU(2)$ instanton gauge fields. Consistently with Milnor's original construction, we employed quaternionic-valued objects for describing the geometric quantities appearing in the metric Ansatz and for performing the necessary calculations to obtain the associated Ricci tensor. Before plugging specific expressions for the gauge fields into the generic formula for the Ricci tensor, we performed a detailed study of the $k=2$ instanton gauge field from the original Ansatz in \cite{PhysRevD.15.1642}, computing its field strength and applying the regularising gauge transformation proposed in \cite{Giambiagi:1977yg}; the above steps are straightforward and well-known for the $k=1$ case. Moreover, we studied the relation between the instantons' moduli space and the Kaluza--Klein metric's moduli space, and found that only a quotient of the former contributes to the latter. Specifically, one should identify all the instantons' configuration which are related via an $SO(5)$ transformation, \ie, via an isometry of the base. This motivated a special choice for the instantons' moduli, which resulted in the corresponding Kaluza--Klein metric having the maximal isometry group: $SO(3) \times O(2)$. It is natural to ask about the possible link between the construction discussed in this paper, for some choice of the $S^4$ radius, and the one proposed by Gromoll and Meyer in \cite{10.2307/1971078}. We were also able to establish a bound on the radius of the base space $S^4$, $r$, which ensures a positive Ricci tensor: $ r > \frac{89}{72}$. 
When the inequality above is met, the $8$-dimensional space-time whose spatial manifold is an exotic sphere satisfies the strong, weak, null and dominant energy conditions. 

From a mathematical point of view, the relatively simple expression for the Riemann tensor, in particular the concrete expressions for the special point in the $k=2$ instanton moduli space, should facilitate an extensive investigation of the behaviour of the sectional curvature. It is known (\cite{nuimeprn10073}) that the Gromoll--Meyer sphere allows metrics with almost everywhere positive sectional curvature. It would be interesting to perform such an investigation for the sectional curvature of the bundle metric at the special points in the $k=1$ and $k=2$ moduli spaces, with the $S^4$ radius $r$ still as a free parameter. Other relevant questions begging for answers concern the geodesic structure of our metric: cut loci and Wiedersehen property, for instance (see \cite{Duran2001}). We leave all these questions for future investigation. Moreover, it would be interesting to repeat the same study for a larger portion of the $k=2$ instanton's moduli space. Generalising it to arbitrary positions is the first step, and including the gauge orientation would exhaust the whole moduli space. At that point, it would be interesting to examine the condition for the metric to be Einstein, and possibly prove a non-existence theorem in case such a condition cannot be satisfied.
Finally, from the physics side, it would be natural to use the explicit results that we derived as a starting point for constructing solutions to supergravity theories in dimensions 7 or higher, supported by appropriate fluxes.

\section*{Acknowledgements}
MC and TSG would like to thank the organisers of ``Mathematical Supergravity" at UNED, Madrid, where their collaboration begun. TSG thanks Leonardo F. Cavenaghi for many valuable suggestions, and for insightful comments on the manuscript. TSG is supported by the Science and Technology Facilities Council (STFC) Consolidated Grants ST/T000686/1 ``Amplitudes, Strings \& Duality'' and ST/X00063X/1 ``Amplitudes, Strings \& Duality''. DSB thanks Pierre Andurand for his generous donation supporting this research.

\appendix

\appendix

\section{Some quaternionic identities}
\label{sec:Appendix_B}
In this section, we collect the quaternionic identities that were used throughout this work. 
A basis for the quaternions $\HH$ is the real $e_0=1$ together with $e_i$, $i=1,2,3$. The multiplication is associative but non-commutative, with
$e_ie_j=-\delta_{ij}+\epsilon_{ijk}e_k$. 
Conjugation is defined by $1\mapsto1$, $e_i\mapsto-e_i$, so with $x=x^ae_a=x^0+x^ie_i$ the conjugated element is 
$\bar x=x^0-x^ie_i$. It satisfies $\overline{xy}=\bar y\bar x$.
The real part, considered as a real number, is $\Re x={1\over2}(x+\bar x)=x^0$, and the imaginary part is
$\Im x={1\over2}(x-\bar x)=x^ie_i$.
A component $x^a$ is extracted from $x\in\HH$ as $x^a=\Re(x\bar e^a)$.
The modulus is defined by $|x|^2=x^ax^a=x\bar x=\bar x x$. It is multiplicative: $|xy|=|x||y|$.
Any non-zero quaternion has a unique inverse $x^{-1}={\bar x\over|x|^2}$.
A useful ``sigma matrix identity" is $x\bar y+y\bar x=2\Re(x\bar y)$.

With the conventions spelled out in Section \ref{sec:'tHooft_notation_and_quaternions}, one finds that the key relation in the dictionary between quaternionic notation and component notation is:
\begin{align}
    \dd x \wedge \dd\bar{x} = - \, ^o\eta^i_{m n}  e_i  \dd x^m \wedge \dd x^n \, ,
\end{align}
where $^o\eta_{i m n}$ are the \textit{reversed} 't Hooft symbols given in \eqref{eq:'tHooft_zero}. Another useful formula comes from considering that a quaternion $x$ left-multiplying the conjugate of another quaternion $\bar{y}$ yields:
\begin{align}
     (x\bar y)_i =- \, ^o\eta_{i m n} x^{m} y^{n} \, ,
\end{align}
where $(\cdot)_i$ stands for the $i^{\mathrm{th}}$ component, and $i=1,2,3$.
A few other identities are easy to see from the fact that the Pauli matrices are traceless. A straightforward one is: $\mathrm{Re}(x) = \frac{1}{2} \mathrm{Tr}(x)$. Moreover, if $x$ is imaginary, then $x= - \mathrm{Re}(x e_i) e_i$.

Moreover, when calculating the field strength from the gauge field (from both $k=1$ and $k=2$ instantons), the following relations become useful:
\begin{align}
    \begin{aligned}
& -2 \operatorname{Re} \dd x \wedge \operatorname{Im} \dd x-\operatorname{Im} \dd x \wedge \operatorname{Im} \dd x=\dd x \wedge \dd\bar{x } \, , \\
& -4 \operatorname{Re} \dd x \wedge \operatorname{Im}\dd x +\dd \bar{x} \wedge \dd x=\dd x \wedge \dd \bar{x} \, .
\end{aligned}
\label{eq:Quat_identity_1_App}
\end{align}
They of course hold for any $\HH$-valued 1-form. Similarly, any identity that holds for matrices in general holds for quaternions, such as $\dd x^{-1} = x^{-1} \dd x x^{-1} $.

A word of caution: elements in $\HH$ (or $\HH'$) are used to encode vectors under the $SO(4)$ of the tangent space of the base $S^4$, but also \eg\ elements in some $\su(2)$ Lie algebra. The index-free notation is efficient, but when one needs the transformation properties, for example when taking a covariant derivative, one needs to keep track of which $\su(2)\oplus\su(2)$ acts by left and right multiplication on the element in question (or $\su(2)$ by commutation, for an element in $\HH'$). Take for example the selfdual part of the Riemann tensor on $S^4$ of eq. \eqref{leftrightS4Riemann}, $R_L={1\over4}E\wedge\bar E$. It is a 2-form taking values in the left $\su(2)_L$ of the $S^4$ tangent space, and fulfills
$D^{(\Omega_L)}R_L=0$ (and in fact even $D_m^{(\Omega_L)}R_L=0$). The maximally symmetric $\su(2)$ 1-instanton field strength is $F={1\over4}E\wedge\bar E$, formally the same expression. Now, however, it is a 2-form valued in the gauge Lie algebra $\su(2)_g$,
and $D^{(A)}_mF=0$.
Consider a more general $\su(2)_g$-valued selfdual 2-form, like $G=\bar u E\wedge\bar Eu$. This is a typical expression for terms in the $k=2$ field strength. Here, $u$ (which is $1$ for the $k=1$ $F$ above) must be thought of as a bifundamental under $\su(2)_L\oplus\su(2)_g$. A covariant derivative of $G$ becomes
\begin{align}
D_mG=D_m\bar uE\wedge\bar Eu+\bar uE\wedge\bar ED_mu\;,
\end{align}
where $D_mu=\*_mu+\Omega_{Lm}u-uA_m$, of course with quaternionic multiplication. For the symmetric 1-instanton, where $u=1$ and ``$\Omega_L=A$'', this vanishes.

\bibliographystyle{utphysmod2}

\bibliography{biblio}

\providecommand{\href}[2]{#2}\begingroup\raggedright\begin{thebibliography}{10}

\bibitem{10.2307/1969983}
J.~Milnor,  {\em On manifolds homeomorphic to the 7-sphere}, Annals of Mathematics {\bf 64}, 399--405 (1956).

\bibitem{YAMAGISHI198447}
K.~Yamagishi,  {\em Supergravity on seven-dimensional homotopy spheres}, Physics Letters B {\bf 134}, 47--50 (1984).

\bibitem{FREUND1985263}
P.~G. Freund,  {\em Higher-dimensional unification}, Physica D: Nonlinear Phenomena {\bf 15}, 263--269 (1985).

\bibitem{Witten:1985xe}
E.~Witten,  {\em {Global gravitational anomalies}}, Commun. Math. Phys. {\bf 100}, 197 (1985).

\bibitem{10.1063/1.529078}
R.~A. Baadhio and P.~Lee,  {\em {On the global gravitational instanton and soliton that are homotopy spheres}}, Journal of Mathematical Physics {\bf 32}, 2869--2874 (10, 1991).

\bibitem{Asselmeyer:1996bh}
T.~Asselmeyer,  {\em {Generation of source terms in general relativity by differential structures}}, Class. Quant. Grav. {\bf 14}, 749--758 (1997) [\href{http://www.arXiv.org/abs/gr-qc/9610009}{{\tt gr-qc/9610009}}].

\bibitem{cavenaghi2024hearingexoticsmoothstructures}
L.~F. Cavenaghi, J.~M. do~Ó and L.~D. Sperança,  {\em Hearing exotic smooth structures}, 2024.

\bibitem{CAVENAGHI2024102121}
L.~F. Cavenaghi and L.~Grama,  {\em {Gromoll–Meyer's actions and the geometry of (exotic) spacetimes}}, Differential Geometry and its Applications {\bf 94}, 102121 (2024).

\bibitem{nuimeprn10073}
M.~Joachim and D.~Wraith,  {\em Exotic spheres and curvature}, Bulletin of the American Mathematical Society {\bf 45}, 595--616 (2008).

\bibitem{10.2307/1971078}
D.~Gromoll and W.~Meyer,  {\em An exotic sphere with nonnegative sectional curvature}, Annals of Mathematics {\bf 100}, 401--406 (1974).

\bibitem{Gherardini:2023uyx}
T.~S. Gherardini,  {\em {Exotic spheres\textquoteright{} metrics and solutions via Kaluza--Klein techniques}}, JHEP {\bf 12}, 100 (2023) [\href{http://www.arXiv.org/abs/2309.01703}{{\tt 2309.01703}}].

\bibitem{Vandoren:2008xg}
S.~Vandoren and P.~van Nieuwenhuizen,  {\em {Lectures on instantons}}, \href{http://www.arXiv.org/abs/0802.1862}{{\tt 0802.1862}}.

\bibitem{BELAVIN197585}
A.~Belavin, A.~Polyakov, A.~Schwartz and Y.~Tyupkin,  {\em {Pseudoparticle solutions of the Yang--Mills equations}}, Physics Letters B {\bf 59}, 85--87 (1975).

\bibitem{tHooft:1976snw}
G.~'t~Hooft,  {\em {Computation of the quantum effects due to a four-dimensional pseudoparticle}}, Phys. Rev. D {\bf 14}, 3432--3450 (1976). [Erratum: Phys.Rev.D 18, 2199 (1978)].

\bibitem{McEnroe2016MILNORSCO}
R.~M. McEnroe,  {\em Milnor’s construction of exotic 7-spheres},
\newblock 2016.

\bibitem{10.1063/1.525753}
R.~Percacci and S.~Randjbar‐Daemi,  {\em {Kaluza–Klein theories on bundles with homogeneous fibers. I}}, Journal of Mathematical Physics {\bf 24}, 807--814 (04, 1983).

\bibitem{DUFF198490}
M.~Duff, B.~Nilsson, C.~Pope and N.~Warner,  {\em {On the consistency of the Kaluza--Klein ansatz}}, Physics Letters B {\bf 149}, 90--94 (1984).

\bibitem{Bailin:1987jd}
D.~Bailin and A.~Love,  {\em {Kaluza--Klein theories}}, Rept. Prog. Phys. {\bf 50}, 1087--1170 (1987).

\bibitem{Salam:1981xd}
A.~Salam and J.~A. Strathdee,  {\em {On Kaluza--Klein theory}}, Annals Phys. {\bf 141}, 316--352 (1982).

\bibitem{ATIYAH1978185}
M.~Atiyah, N.~Hitchin, V.~Drinfeld and Y.~Manin,  {\em Construction of instantons}, Physics Letters A {\bf 65}, 185--187 (1978).

\bibitem{PhysRevD.15.1642}
R.~Jackiw, C.~Nohl and C.~Rebbi,  {\em Conformal properties of pseudoparticle configurations}, Phys. Rev. D {\bf 15}, 1642--1646 (Mar, 1977).

\bibitem{Giambiagi:1977yg}
J.~J. Giambiagi and K.~D. Rothe,  {\em {Regular $N$-instanton fields and singular gauge transformations}}, Nucl. Phys. B {\bf 129}, 111--124 (1977).

\bibitem{Curiel:2014zba}
E.~Curiel,  {\em {A primer on energy conditions}}, Einstein Stud. {\bf 13}, 43--104 (2017) [\href{http://www.arXiv.org/abs/1405.0403}{{\tt 1405.0403}}].

\bibitem{Martin-Moruno:2017exc}
P.~Martin-Moruno and M.~Visser,  {\em {Classical and semi-classical energy conditions}}, Fundam. Theor. Phys. {\bf 189}, 193--213 (2017) [\href{http://www.arXiv.org/abs/1702.05915}{{\tt 1702.05915}}].

\bibitem{Econditions}
D.~Allison,  {\em {Energy conditions in standard static spacetimes}}, General Relativity and Gravitation {\bf 20} (1988).

\bibitem{Duran2001}
C.~E. Durán,  {\em Pointed wiedersehen metrics on exotic spheres and diffeomorphisms of s6}, Geometriae Dedicata {\bf 88}, 199--210 (2001).

\end{thebibliography}\endgroup

\end{document}